\documentclass[a4paper,12pt]{article}

\usepackage{amsmath}\usepackage{amssymb}
\usepackage{latexsym}\usepackage{cite}

\usepackage[dvips]{graphicx}
\usepackage{pstricks}
\usepackage{color}
\usepackage{bm}

\usepackage{graphicx}

\usepackage{amsmath, amssymb, graphics}

 \definecolor{red}{rgb}{1,0,0} 
 \definecolor{blue}{rgb}{0,0,1} 
 \definecolor{green}{rgb}{0,1,0} 
 \definecolor{bgreen}{rgb}{0,0.5,0.5} 
 \definecolor{magenta}{rgb}{1,0,1} 

\newcommand{\mathsym}[1]{{}}

\newcommand{\be}{\begin{equation}}
\newcommand{\ee}{\end{equation}}

\newcommand{\ka}{\kappa}

\def\beq{\begin{equation}}
\def\eeq{\end{equation}}
\def\beqr{\begin{eqnarray}}
\def\eeqr{\end{eqnarray}}
\def\pl{\partial}

\def\al{\alpha}
\def\bt{\beta}
\def\Ga{\Gamma}
\def\ga{\gamma}
\def\de{\delta}
\def\De{\Delta}

\def\ka{\kappa}
\def\si{\sigma}

\def\te{\theta}

\def\La{\Lambda}
\def\lam{\lambda}

\def\om{\omega}
\def\ep{\epsilon}

\def\vp{\varphi}

\def\sq{\sqrt}

\def\l{\left (}
\def\r{\right )}

\def\fr{\frac}
\def\la{\label}
\def\hs{\hspace}
\def\vs{\vspace}

\def\ran{\rangle}
\def\lan{\langle}
\def\ov{\overline}
\def\tl{\tilde}
\def\tm{\times}

\begin{document}

\begin{flushright}
December 31, 2025 \\
\end{flushright}

\vs{1.5cm}

\begin{center}
{\Large\bf

Minimal Modular Flavor Symmetry and

Lepton Textures Near Fixed Points}

\end{center}

\vspace{0.5cm}
\begin{center}
{\large
{}~Zurab Tavartkiladze\footnote{E-mail: zurab.tavartkiladze@gmail.com}
}
\vspace{0.5cm}

{\em Center for Elementary Particle Physics, ITP, Ilia State University, 0179 Tbilisi, Georgia}

\end{center}
\vspace{0.3cm}

\begin{abstract}


An extension of the Standard Model with $\Gamma_2\simeq S_3$ modular flavor symmetry is presented.
We consider the construction of the lepton sector, augmented by two right-handed neutrino states,
in the vicinity of the fixed points $\tau = i\infty $, $\tau = i$ 
and $\tau \!=\!\omega \!=\!-\frac{1}{2}\!+\!i\frac{\sqrt{3}}{2}$. Due to the residual symmetries
at these points, and with the aid of nonholomorphic modular forms (which constitute representations
of $S_3$) and by assigning specific transformation properties to the fermion fields, highly
economical models (without flavon fields) are constructed with interesting Yukawa textures.
All presented models strongly prefer  the inverted ordering for the neutrino masses.

\end{abstract}

\hspace{0.4cm}{\it Keywords:}~Modular flavor symmetry;  Lepton masses; Neutrino oscillations.

\hspace{0.4cm}PACS numbers: ~11.30.Hv, 12.15.Ff, 14.60.Pq





\section{Introduction}

The hierarchies between quark and lepton masses and the suppressed values of the Cabibbo-Kobayashi-Maskawa
 matrix elements are unexplained within the
Standard Model (SM). Besides these, the neutrino data \cite{Capozzi:2021fjo}  cannot be accommodated within the SM. The latter already indicate the
presence of new physics, and the simplest extension is to add right-handed neutrino (RHN) states. On the other hand, for understanding
the flavor puzzle, one can postulate some flavor symmetry. The simplest version of this is the Abelian $U(1)_F$ flavor symmetry \cite{Froggatt:1978nt},
 which is desirable to gauge. Although there has been quite a bit of success in building models with gauged flavor $U(1)_F$
 \cite{Dudas:1995yu,Chen:2008tc,Tavartkiladze:2011ex}, this approach is somewhat challenging
 because additional constraints from the requirement of anomaly cancellation need to be imposed.

 Recently, models based on modular flavor symmetries have attracted considerable attention. This direction was initiated by a
pioneering paper \cite{Feruglio:2017spp}, after which the number of works in this area began to grow rapidly- a trend that
continues to this day
\cite{Feruglio:2017spp,Feruglio:2021dte,Kobayashi:2021pav,Kikuchi:2023cap,Kobayashi:2023zzc,
Feruglio:2023mii,Meloni:2023aru,Marciano:2024nwm,Nomura:2024ouj,Du:2020ylx,Kobayashi:2018vbk,
Kobayashi:2019rzp,Granelli:2025lds,Qu:2025ddz,Kumar:2023moh,Novichkov:2018ovf,
Ding:2019gof,Okada:2020ukr,Qu:2024rns}.
 These constructions allow building simple models with modular finite non-Abelian flavor symmetries
 with a very economical scalar sector involving a single complex modulus field $\tau $.
 With the same $\tau$, the modular states are constructed, generating the effective Yukawa couplings and also responsible for
 the high flavor-symmetry breaking without introducing additional flavon states. The smallest modular finite group is
  $\Ga_2 \simeq S_3$. As a flavor symmetry, $S_3$ has a long history dating back to the pioneering work of Ref. 
  \cite{Pakvasa:1977in}. However,
  constructions with the modular flavor $S_3$ symmetry have gained new insights \cite{Meloni:2023aru, Marciano:2024nwm,Nomura:2024ouj,Du:2020ylx,Kobayashi:2018vbk,Kobayashi:2019rzp}, and because of its simplicity we aim to investigate
  it in the present work. In this paper, we focus only on the lepton sector, which will be enlarged by the RHNs.

In the next section we discuss some properties of the $\Ga_2 \simeq S_3$ modular symmetry and the construction of nonholomorphic
modular forms, which will belong to representations of $S_3$. In Secs. \ref{sec-Fix-inf}, \ref{sec-Fix-i},
and \ref{sec-Fix-om} we present models near
the $\tau = i\infty$, $\tau = i$  and $\tau \!=\!\omega \!=\!-\frac{1}{2}\!+\!i\frac{\sqrt{3}}{2}$
 fixed points, respectively. As we will see, the constructions considered can be very
economical and allow for predictions. Specifically, due to residual discrete symmetries at these fixed points, all
considered neutrino scenarios strongly suggest the inverted ordering of the neutrino masses.
The paper includes three appendices. In Appendix \ref{app-S3}, some properties of the $\Ga_2 \simeq S_3$ symmetry, representations of
nonholomorphic modular forms, and various useful expressions are derived. Also, the structures of modular representations
 at the considered fixed points are given. In Appendix \ref{app-kin}, the invariant kinetic couplings are presented, and simple ways
  of fixing the modulus field, with desirable values, are discussed. Appendix \ref{ApU} provides the neutrino parametrization
   and some relations between the observables and the parameters of the specific model we are proposing.

\section{SM with $\Ga_2\simeq S_3$ Modular Symmetry}

The model we propose is pretty simple. It extends the SM by incorporating right-handed
 neutrinos $N_i$ (described below) along with the complex field $\tau $,
\beq
\tau=x+iy ~,
\la{tau}
\eeq
where both components $x$ and $y$ are functions of the four spacetime coordinates $x_\mu$.
The field $\tau $ is central to the flavor modular symmetry, under which $\tau $ transforms as follows:
\beq
\tau'=\fr{a\tau +b}{c\tau +d}~,~~~~~{\rm with }~~ad-cd=1,~~ \{a,b,c,d\}\in  \mathbb{Z} .
\la{tau-to-tau1}
\eeq
The $\tau $ gets values at the upper half of the complex plane. The transformation (\ref{tau-to-tau1}) allows us
to consider the values of  $\tau $ picked from the fundamental domain 
\beq
{\cal F}=  \left\{|\tau |\geq 1, ~-\fr{1}{2}\leq x< \fr{1}{2}, ~y>0   \right\} .
\la{fun-dom}
\eeq
Note that points with $Re(\tau )\!=\!x\!=\!\!1/2$ are excluded from the fundamental domain
${\cal F}$ because the $T$ transformation $\tau \to \tau+1$ [realized with $a\!=\!b\!=\!d\!=\!1, c\!=\!0$ in (\ref{tau-to-tau1})] maps points with $Re(\tau) =- 1/2$  onto the points with $Re(\tau) = 1/2$.

The boundary of ${\cal F}$ includes the fixed points,
\beq
{\rm Fixed ~points}:~~~\tau =\{i\infty,~ i, ~\om \!=\!-\fr{1}{2}\!+\!i\fr{\sqrt{3}}{2}\}.
\la{fixed-pnts}
\eeq
[The fixed point $1/2+i\fr{\sqrt{3}}{2}$ is excluded from (\ref{fixed-pnts}) 
because it is related to $\omega $ via the $T$ transformation, as noted following Eq. (\ref{fun-dom})].

The transformation properties of the fermionic and Higgs doublet fields will be addressed later.
We begin our discussion with the modular forms, which are not independent fields but rather functions of the field $\tau $.

The modular form $f^{(k)}$ of weight $k$ is a function of $\tau $ with the following transformation property:
\beq
f^{(k)}\l \fr{a\tau +b}{c\tau +d}\r =(c\tau +d)^kf^{(k)}(\tau )~.
\la{fk}
\eeq
If one restricts attention to holomorphic forms, without further constraints on the integer numbers $\{a, b, c, d\}$
[from Eq. (\ref{tau-to-tau1})], the basis for the holomorphic even-weight forms is formed by the weight $4$ and $6$
forms $E_4(\tau )$ and    $E_6(\tau )$, respectively  \cite{apostol, zagier}.
Thus, the holomorphic $2p$ weight form can be expressed as the superposition $\sum_{m,n}c_{mn}E_4^mE_6^n$, where the non-negative
integers $(m,n)$ are all possible solutions to the equation  $4m+6n=2p$. The modular forms $E_{4,6}$ and their expansions
are provided in Appendix \ref{app-S3} [see Eq. (\ref{E4-E6})].

In this paper, we will focus on general (i.e., nonholomorphic) even-weighted modular forms.
For their building, the minimal positive even-weighted
  form $\tl E_2$ (with $k=2$) will be used.  The $\tl E_2$ is nonholomorphic and given in Eq. (\ref{E2tl}). This form is constructed using the Eisenstein
  series $E_2$  given in Eq. (\ref{E2}).
Besides this, the factor $ \ka =(i\tau-i\bar{\tau})/2$ will be used, which
under the modular transformation (\ref{tau-to-tau1}) transforms as:
\beq
   \ka =\fr{1}{2}(i\tau-i\bar{\tau})\to \fr{1}{2}\fr{(i\tau-i\bar{\tau})}{(c\tau +d)(c\bar{\tau}+d)} .
\la{kapa-transf}
\eeq
With the $\ka $,  $\tl E_2$, and $E_{4,6}$ (and their conjugates) the modular form $f^{(k)}$ can be constructed:
$$
f^{(k)}(\tau )=\sum C_{lmnp}^{\ov l\ov m \ov n}
 \ka^{2p}\tl E_2^{\hs{0.03cm} l}E_4^{\hs{0.03cm} m}E_6^{\hs{0.03cm} n} (\tl E_2^{\hs{0.05cm}\ov l}E_4^{\hs{0.05cm}\ov m}E_6^{\hs{0.05cm}\ov n})^*  ,
$$
\beq
{\rm with}~~~k=2(l+2m+3n-p),~~~~{\ov l}+2{\ov m}+3{\ov n}=p~ ,
\la{fk-from-kaEs}
\eeq
which has a transformation property given in (\ref{fk}).
Restricting to holomorphic forms, only even-weight forms with $k\geq 4$  are nonzero \cite{apostol, zagier}. 
Note that nonsupersymmetric constructions do not require holomorphicity,\footnote{In the literature, the terms 'holomorphic forms' 
and 'entire modular forms' are used interchangeably.  For the form $\tl E_2$ and the combinations in (\ref{fk-from-kaEs}), which involve 
different powers of $\ka $, the term  'almost modular' is employed. For strict definitions see \cite{kaneko, apostol, zagier,Nagatomo}.}
 so we should also consider nonholomorphic forms - such as $\tl E_2$ and the combinations in Eq.  (\ref{fk-from-kaEs}) - with arbitrary even weights
 (including negative ones) allowing positive powers of $\tl E_2$, $E_{4,6}$, and $\kappa$.
 During model building, when a large value of $|\lan \kappa \ran|=\lan y\ran$ is required, the physical requirement is to use 
 the low powers of $\ka $ in (\ref{fk-from-kaEs})  to avoid large effective coupling. On the other hand, with $\lan y\ran \sim 1$, no additional restriction
is required for the construction of $f^{(k)}$. For the same reason, it is common not to use inverse powers of modular 
forms [also in supersymmetric (SUSY) constructions] to avoid poles or large values at cusp points or in their vicinity.

One may also apply the weight-raising and weight-lowering operators \cite{Nagatomo} - $\hat {\bf \pl }$  and $\hat {\bf \de^*}$ 
 [see Eq. (\ref{del-de-ops}) and the discussion therein]. When acting on a modular form of weight $k$, $\hat {\bf \pl }$  and $\hat {\bf \de^*}$  
 yield modular forms of weights $k+$2 and $k-2$, respectively. In Appendix A we present simple examples to illustrate that
  the modular forms obtained in this way are contained in (\ref{fk-from-kaEs}).

Denoted by $\Ga $, the transformations (\ref{tau-to-tau1}) and (\ref{fk}) (without any additional constraints on integers $\{a, b, c, d \}$) form a group.
 In conjunction with $\Ga $ one can also consider the group $\Ga (N)$ ($N=1,2,\cdots $),
  \beq
  \Ga (N):\tau'=\fr{a\tau +b}{c\tau +d}~,~~ad-cd\!=\!1,~~ \{a,b,c,d\}\!\!\in  \mathbb{Z}~~{\rm and}~~
  \left(
  \begin{array}{cc}
   a & b \\
   c & d \\
    \end{array}
     \right)\!=\!I_{2\tm 2}~ ({\rm mod}~ N).
  \la{GamN}
  \eeq
While $\Ga $ and $\Ga (N)$ form infinite groups, the elements of the quotient group $\Ga_N=\bar \Ga /\bar \Ga (N)$
  [where $\bar \Ga =\Ga/\{\pm 1\}$ and $\bar \Ga (N)=\Ga (N)/\{\pm 1\}$] are a finite modular group of level $N$ \cite{Feruglio:2017spp}.
Consequently, acting on the modular forms, the $\Ga_N$  transform them as representations of the corresponding finite group
\cite{deAdelhartToorop:2011re},\cite{Feruglio:2017spp}.

 The $\Ga_N $ groups, along with the singlets, possess nontrivial representations that enable the construction
 of compelling and intriguing modular flavor symmetric models
 \cite{Feruglio:2017spp,Feruglio:2021dte,Kobayashi:2021pav,Kikuchi:2023cap,Kobayashi:2023zzc,
 Feruglio:2023mii,Meloni:2023aru,Marciano:2024nwm,Nomura:2024ouj,Du:2020ylx,Kobayashi:2018vbk,
 Kobayashi:2019rzp,Granelli:2025lds,Qu:2025ddz,Kumar:2023moh,Novichkov:2018ovf,
 Ding:2019gof,Okada:2020ukr,Qu:2024rns}.
 Since the simplest and smallest case is $\Ga_2\simeq S_3$, this work will focus on this option and explore it in detail.
 Specifically, we will examine the lepton sector and demonstrate that, under certain conditions,
 the models we construct can be as successful and predictive as supersymmetric constructions.

The modular group $\Ga_2\simeq S_3$ is a finite non-Abelian group   with six
elements $g_i$ (where $i=1,\cdots \!,6$). This group possesses, in addition to the singlet
representation ${\bf 1}$, a pseudosinglet representation ${\bf 1'}$  and a doublet representation
${\bf 2}$.

The modular forms $\tl E_2, E_{4,6}$, and  $f^{(k)}(\tau )$ [given in Eq. (\ref{fk-from-kaEs})],  are singlets under $S_3$. However, from these forms, one can
construct the weight $k$ doublet $D^{(k)}=(D^{(k)}_1, D^{(k)}_2)\sim {\bf 2}$  using the expressions in Eq.  (\ref{Dk}) 
(for details, see Appendix \ref{app-S3}).
One can also multiply modular forms belonging to  different representations of $S_3$, and, by using
the multiplication rules (e.g., ${\bf 2}\tm {\bf 2}={\bf 2}+{\bf 1}+{\bf 1'}$) given in (\ref{2by2})-(\ref{1by1pr}), construct
modular forms in the required representations with the desired weights.

Thus far, we have discussed modular forms that are functions of the single field $\tau $
and possess definite transformation properties under the modular group. However, the SM states and the RHNs are not modular forms,
nonetheless, they will exhibit well-defined transformation properties under the modular group $\Ga_2\simeq S_3$.

In particular, when the transformations (\ref{fk}) and (\ref{GamN}) are  applied, the fermions $\psi $ will transform as follows:
\beq
\psi '=(c\tau+d)^{k_{\psi  }}\rho \cdot \psi ~,
\la{ferm-transf}
\eeq
where the operator $\rho $ corresponds to the representation of $\psi $. If the latter is the 
$S_{3}$ doublet, then $\rho $ is a $2\tm 2$
matrix representing the group element in the doublet representation.
Two generating elements, $\rho (T)$ and $\rho (S)$, in the doublet representation are given in (\ref{rho2-TS}).
The remaining elements are expressed in terms of these via the combinations shown in (\ref{S3-elements}),
following the rules outlined in Eq. (\ref{Dk-transf}). Clearly, if $\psi $ is in the ${\bf 1}$ or ${\bf 1'}$
representation, then $\rho =1$ or $\rho =-1$, respectively.
The  $k_{\psi }$  in (\ref{ferm-transf}) represents the corresponding weight. Here and henceforth,
for fermions  $\psi $, we will assume the use of a two-component Weyl spinor.

Similarly, the SM Higgs doublet $\varphi $ transforms as  follows:
\beq
\varphi'=\pm (c\tau +d)^{k_{\varphi }}\varphi ~.
\la{Higs-trans}
\eeq
The ``$+$" sign in (\ref{Higs-trans}) applies when $\varphi \sim {\bf 1}$, whereas if 
$\varphi $  is a pseudosinglet, $\varphi \sim {\bf 1'}$, the sign should be ``$-$".

The structure of the scalar potential and Yukawa interactions depends on the assignment
of representations and weights, which will be discussed in the following sections.
The invariant kinetic terms for states with  specific weights are given in Appendix B.

We would like to emphasize that the modular forms are not independent; rather, they 
are functions of $\tau $ and can simultaneously mimic the nontrivial representations of 
$S_3$. Therefore, to achieve the breaking 
of the $S_3$, we do not need to introduce additional flavon states. As a result, the extension 
of the SM's scalar sector is limited to the field $\tau $. In the fermion sector, we will introduce 
RHN states alongside the SM fermions, as RHNs are vital for the generation of neutrino masses.
Consequently, the proposed extension - based on the $SU(3)_c\tm SU(2)_L\tm U(1)_Y\tm S_3$ 
symmetry - can be regarded as a minimal extension with non-Abelian symmetry.

Although the extension we are considering is relatively economical, in general,
nonsupersymmetric constructions that permit nonholomorphic operators typically lead to an 
increase in couplings.

Among the holomorphic modular forms,  the weight-two doublet is $Y = (Y_1, Y_2)$.
In addition to this, there exists another independent doublet, the weight-four doublet 
$Y^{(4)}$, defined as $Y^{(4)} \propto  (Y\!\! \cdot \!\!Y)_{\mathbf{2}} = (Y_2^2 - Y_1^2, 2Y_1Y_2)$. 
No other independent holomorphic doublets exist. Doublets of higher weights can be expressed 
in terms of $Y$, $Y^{(4)}$, and the holomorphic functions $E_4$ and $E_6$.
Since we are not considering supersymmetry, holomorphy is not necessary, and we will have 
five independent doublets, given in (\ref{5-dublets}). From the latter one can build the five 
independent doublets, of any fixed even weight $k$, which are given as
 \begin{align}
 {\rm doublets~of~ weight}~k, ~ D^{(k)}=\left \{ \right. &  Yf^{(k-2)}(\tau ),~~Y^*\ka^2f^{(k+2)}(\tau ), ~~
 (Y\!\!\cdot \!\!Y)_{\bf 2}f^{(k-4)}(\tau ),\nonumber \\
       & (Y\!\!\cdot \!\!Y^*)_{\bf 2}\ka^2f^{(k)}(\tau ),  ~~ (Y^*\!\!\cdot \!\!Y^*)_{\bf 2}\ka^4f^{(k+4)}(\tau ) \left.  \right \} 
        \la{nonhol-Ds}
\end{align}
(see Appendix \ref{app-S3} for a discussion),
where $f(\tau)$'s are the singlet modular forms, with corresponding weights,  given by Eq. (\ref{fk-from-kaEs}).
As an illustrative example, the five doublets of weight $k=2$ are constructed in (\ref{k2-5doubl}) using \eqref{nonhol-Ds}.

Through the modular doublets of different weights, using the multiplication rule (\ref{2by2}),  the
pseudosinglet modular form with weight $k$, can be constructed,
\beq
{\bf(1')}^{(k)}=\l D^{(k-n)}\cdot D^{(n)}\r_{1'} =D_1^{(k-n)}\cdot D_2^{(n)}-D_2^{(k-n)}\cdot D_1^{(n)}~.
\la{1prime-k}
\eeq

Having numerous independent doublets will lead
 to an increase in the number of invariants and unknown couplings, which in turn reduces the 
 predictive power of the construction.
However, there is a solution to this unpleasant situation.
As is well known, the fundamental domain of the $\Ga_2\simeq S_3$ includes so-called fixed points, 
where enhanced symmetries - specifically the discrete symmetries, which are subgroups of 
$\Ga_2\simeq S_3$ - are present. At these fixed points, either some modular forms vanish or the doublets 
become aligned. This can significantly reduce the number of parameters, and as we will demonstrate, 
the model can be as predictive as supersymmetric constructions. In the following section, we will 
discuss the lepton sector in the vicinity of the $\tau =i\infty $ fixed point,
while in Secs. \ref{sec-Fix-i} and  \ref{sec-Fix-om}  the models near the $\tau =i$
and $\tau \!=\!\omega $
 fixed points, respectively, are presented.

\section{Model Near $\tau =i\infty $ Fixed Point: ``Intermediate" $Z_2^T$  Symmetry}
\label{sec-Fix-inf}

As discussed in Appendix \ref{ap-Rep-iinf},
at $\tau=i\infty $ fixed point the $S_3$ doublet modular form of any weight $D^{(k)}$ has the second component $D^{(k)}_2=0$, 
and also any pseudosinglet
$({\bf 1'})^{(k)}$ modular form vanishes:
\beq
{\rm at}~~\tau=i\infty:~~~~~~~~D^{(k)} = \left(\!\!
      \begin{array}{c}
        1 \\
        0 \\
      \end{array}\!\!\right)~,~~~({\bf 1'})^{(k)}=0 ,
\la{D1pr-inft}
\eeq
where for $D^{(k)}$ we are omitting the prefactor, bearing in mind that the latter can be absorbed in the appropriate
coupling constant of the Lagrangian.
Thus, at $\tau=i\infty $ the $Z_2^T$ symmetry is unbroken, because action of $\rho_{[2]}(T)$ on $D^{(k)}(\tau)$ renders it,
and also any singlet modular form,
invariant:
\beq
{\rm at}~\tau=i\infty ,~~~Z_2^T:~~~~~~~D^{(k)}(\tau+1) =\rho_{[2]}(T)D^{(k)}(\tau ),~~~({\bf 1})^{(k)}\to  ({\bf 1})^{(k)},
\la{D2pr-inft}
\eeq
where the $\rho_{[2]}(T)$ is given in (\ref{rho2-TS}).
On the other hand the fields in the ${\bf 2}$, ${\bf 1}$ and ${\bf 1'}$ representations of $S_3$, under the $Z_2^T$ (which is subgroup of $S_3$),
transform as
\beq
Z_2^T:~~~
\left(\!\!
      \begin{array}{c}
        \psi_1 \\
        \psi_2 \\
      \end{array}\!\!\right)\to \left(\!\!\begin{array}{c}
        ~\psi_1 \\
        \!\!-\psi_2 \\
      \end{array}\!\!\right),~~~~{\bf 1}\to {\bf 1},~~~~ {\bf 1'}\to -{\bf 1'}.
\la{fields-Z2T}
\eeq
Thus, any texture zero that is obtained due to $Z_2^T$ will be protected by the same symmetry.\footnote{Residual discrete symmetries, i.e., subgroups of larger finite symmetries, can significantly aid in preserving
texture zeros at the required level \cite{Feruglio:2021dte,Feruglio:2023mii,Kikuchi:2023cap,Kobayashi:2023zzc,Kobayashi:2021pav,Granelli:2025lds,deMedeirosVarzielas:2025byb}.}

With mild violation of $Z_2^T$, the modular forms slightly deviate from their original values. In practice, instead of $Im(\tau )=y\to \infty $
it is sufficient to have $y\stackrel{>}{_\sim }2$ because the small parameter, which characterizes the $Z_2^T$ symmetry breaking effects is $\ep = e^{-\pi y}\ll 1$.
Therefore, for the doublets and the pseudosinglets, we can use
\beq
{\rm with}~~ Im(\tau ) \stackrel{>}{_\sim }2:~~~
D^{(k)} = \left(\!\!
      \begin{array}{c}
        1 \\
        \ep \\
      \end{array}\!\!\right)~,~~~({\bf 1'})^{(k)}\sim \ep ,
\la{D1pr-y}
\eeq
where for the $\ep $ parameter ($|\ep|\ll 1$) appearing in the doublets with expressions of Eq. (\ref{DevsD1})
 we have (see Appendix \ref{ap-Rep-iinf} for an explanation)
\beq
|\ep|\sim 8\sqrt{3}e^{-\pi y}~~~~{\rm or}~~~~16\sqrt{3}e^{-\pi y}.
\la{eps-val}
\eeq
Thus,  the $\ep $ will serve as a small expansion parameter.
In Appendix \ref{app-kin} we discuss and give the potential, which ensures to fix the needed value of $\lan \tau \ran$ near the fixed point
$\tau =i\infty $.

\subsection{Charged Lepton Sector}
\label{ME-inf}

Since we are focusing on the lepton sector, we begin by outlining the transformation properties of the lepton doublets
$l_{1,2,3}=(\nu, e^{-})_{1,2,3}$, the iso-singlet charged leptons $e^c_{1,2,3}$, and the Higgs field $\varphi$.
The RHNs will be discussed subsequently, when constructing the neutrino sector.
The leptons $l_1$ and $l_2$ will be embedded in the $S_3$ doublet $L=(l_1, l_2)^T$, while the remaining leptons will be $S_3$ pseudosinglets. The Higgs doublet will be the singlet of $S_3$. Therefore, these states are the following representations of the group $S_3$:
\beq
L=\left(\!\!
      \begin{array}{c}
        l_1 \\
        l_2 \\
      \end{array}\!\!\right)\sim {\bf 2},~~~~l_3\sim {\bf 1'},~~~~ e^c_{1,2,3}\sim {\bf 1'},
      ~~~~\varphi \sim {\bf 1}.
\la{S3-reps}
\eeq

With the transformations given in (\ref{S3-reps}) and by the following weight assignments
\beq
k_{L}=-2,~~~k_{l_3}=k_{e^c_i}=k_{\varphi }=0,
\la{le-weights}
\eeq
without loss of any generality, Lagrangian couplings relevant for the charged lepton masses and are invariant under all symmetries, are
\beq
-{\cal L}_E=\ga_0\tl{\varphi }l_3e^c_3-\tl{\varphi }(LD^{(2-m)})_{1'}{\bf 1}^{(m)}(\bt_0e^c_2+\bt_0'e^c_3)
+\tl{\varphi }(LD^{(2-m)})_{1}({\bf 1'})^{(m)}(\al_0e^c_1+\al_0'e^c_2+\al_0''e^c_3) .
\la{ll-yuk}
\eeq
Upon writing the Yukawa coupling in (\ref{ll-yuk}), we have  chosen the basis in which $e^c_3$ couples only with $l_3$.
In the remaining terms, we have displayed only the relevant invariants.

Using in (\ref{ll-yuk}) the structure of doublets, singlets, and pseudosinglets near the $\tau=i\infty $ fixed point [see Eq. (\ref{D1pr-y})]
and using canonical normalization of the states (details discussed in Appendix \ref{app-kin}), the charged lepton mass matrix
in the $(l_1, l_2, l_3)M_E(e^c_1, e^c_2, e^c_3)^T$ basis  will be:
\beq
M_E=\!\left(
      \begin{array}{ccc}
        \al \ep_e & \al'\ep_e  & \al'' \ep_e \\
       0  & \bt & \bt' \\
       0 & 0 & \ga \\
      \end{array}
    \right)\!\!v +{\cal O}(\ep_e^2),
\la{ME}
\eeq
where $\ep_e\sim \ep $. 
From
the second term of Eq.(\ref{ll-yuk}),  the couplings $\sim v\ep_el_1(\bt_0e^c_2+\bt_0'e^c_3)$ arise, 
which are included in the first row of matrix (\ref{ME})  [the constants $\al'$ and $\al''$
 are linear combinations of the original couplings $(\al_0', \bt_0)$ and $(\al_0'', \bt_0')$, respectively]. 
The properly suppressed $\ep \sim e^{-\pi y} $ can  be naturally obtained with $y>1$.
 We see that in the limit $\ep_e\to 0$ [i.e. with unbroken $Z_2^T$, which means configurations 
 of Eq. (\ref{D1pr-inft})] only $\tau $ and $\mu $ leptons gain masses, while the electron remains massless.
 While in this setting one cannot explain the small value of the $\lam_{\tau }(\sim 10^{-2})$ and also  the  hierarchy
 $m_{\mu }/m_{\tau}(\approx 1/17)$, it is still
 satisfactory that $m_e$ is generated due to the $Z_2^T$ symmetry breaking and suppressed with the small 
 parameter $\ep_e\sim \lam_e\sim 10^{-6}$.
 Since $S_3$'s largest representation
 is the doublet and there is only $Z_2^T$ 'intermediate' symmetry, only one small parameter $\ep_e$ is appearing within this setup.
  Perhaps, within larger finite modular groups, such as $A_4, S_4$, etc.,  it is worthwhile to attempt \cite{in-prep}
 to explain the suppressed values of  $\lam_{\tau }$  and of the ratio $m_{\mu}/m_{\tau}$, in the spirit of Refs.
 \cite{Kobayashi:2021pav,Kikuchi:2023cap,Kobayashi:2023zzc,Granelli:2025lds, Qu:2025ddz}.

By the assumption
\beq
\bt, \bt'  \sim \fr{\ga }{10},~~~
 \bt, \bt'\gg \ep_e,
\la{ga-bt-ep}
\eeq
 we can obtain
\beq
m_{\tau }\simeq |\ga |v ,~~~~ m_{\mu }\simeq |\bt |v  , ~~m_e\simeq |\al \ep_e|v~.
\la{mass-e-mu-tau}
\eeq
Diagonalization of $M_E$ is achieved  by the biunitary transformation
$$
U_lM_EU_{e^c}=M_E^{\rm Diag},
$$
\beq
{\rm with}~~~U_l\!\simeq
                                     \!\! \left(  \begin{array}{ccc}
                                              1 & -\fr{\al'\ep_e}{\bt }  & 0 \\
                                              \!\l \!\fr{\al'\ep_e}{\bt } \! \r^{\!\!*}  & 1 & 0\\
                                              0 & 0 & 1 \\
                                            \end{array}
                                          \right)
                                      \!\! \left(  \begin{array}{ccc}
                                              1 & 0 & -\fr{\al''\ep_e}{\ga }  \\
                                              0 & 1 & 0\\
                                              \!\l \!\fr{\al''\ep_e}{\ga }\! \r^{\!\!*}  & 0 & 1 \\
                                            \end{array}
                                          \right)
                                          \!\!\left(
                                            \begin{array}{ccc}
                                              1 & 0 & 0 \\
                                              0 & 1 & -\fr{\bt'}{\ga } \\
                                              0 & \l\fr{\bt'}{\ga }\r^{\!\!*}  & 1 \\
                                            \end{array}
                                          \right).
\la{mE-diag}
\eeq
This result, obtained upon the assumption (\ref{ga-bt-ep}), corresponds to the hierarchical structure of the mass matrix $M_E$ (\ref{ME}).

\subsection{Neutrino Sector}
\label{sec-inf-nu}

%
%

To implement the concept discussed in the previous section, we will investigate the generation of neutrino
masses through the type-I seesaw mechanism. To do this, we introduce two right-handed neutrinos, $N$ and $N'$,
 which are associated with the representations ${\bf 1}$ and  ${\bf 1'}$ (respectively) of $\Ga_2\simeq S_3$,
 \beq
 S_3:~~~~~~~~N\sim {\bf 1}~,~~~~~N\sim {\bf 1'}~.
 \la{N-transf}
 \eeq
With this extension and the weight assignments
\beq
k_{N}=k_{N'}=0,
\la{k-N-inft}
\eeq
the relevant Lagrangian terms are
$$
-{\cal L}_{\nu }=\left [ (LD^{(n+2)})_1N{\bf 1}^{(-n)} + (LD^{(n+2)})_{\bf 1'}N'{\bf 1}^{(-n)} +l_3N'\right ]\vp
 $$
 \beq
 -\fr{1}{2}\left [ MNN +nMN'N'+2M'NN'({\bf 1'})^{(0)}\right ],
 \la{lN-NN}
 \eeq
where in the Dirac-type couplings we have not included terms that would give the corrections of the order of$\sim \ep_e$.
Thus, in the $\tau \to i\infty $ limit we have
\beq
m_D^{(0)}=\!\left(
      \begin{array}{cc}
        b_0 & 0 \\
        0 & a_0  \\
       0 & 1 \\
      \end{array}
    \right)\!\! \lam v ,~~~
M_N^{(0)}=\!\left(
      \begin{array}{cc}
        1 & 0 \\
        0 & n  \\
      \end{array}
    \right)\!\! M ,
\la{mD-MN-00}
\eeq
where the bases $(\nu_e, \nu_{\mu }, \nu_{\tau})m_D^{(0)}(N, N')^T$ and
$\fr{1}{2}(N, N')M_N^{(0)}(N, N')^T$ have been used.
Structures in (\ref{mD-MN-00}) exhibit residual  $Z_2^T$ symmetry $\{l_1, N\}\!\to \!\{l_1, N\}, \{l_2, l_3, N'\}\!\to \!-\{l_2, l_3, N'\}$. When incorporating contributions that break this $Z_2^T$ symmetry, we can safely ignore changes in $m_D^{(0)}$
 because with $\al \sim (1/3-3), \ga \simeq \lam_{\tau }\sim 10^{-2}$, the parameter $\ep_e $ [see Eq. (\ref{mass-e-mu-tau})],
  \beq
 \ep_e\simeq \fr{\ga }{\al }\fr{m_e}{m_{\tau}}=(0.1-1)\cdot 10^{-5},
  \la{est-epe}
  \eeq
is sufficiently small. Conversely, due to the distinct nature of the Majorana mass terms related to their origins, the effects of  $Z_2^T$ breaking can be more significant in $M_N$. Therefore, for the neutrino
 Dirac and Majorana matrices we will consider the following forms:
\beq
m_D=U_lm_D^{(0)}~ ,~~
M_N=\!\left(
      \begin{array}{cc}
        1 & -\ep_0 \\
        -\ep_0 & n  \\
      \end{array}
    \right)\!\! M ,
    \la{mD-MN-0}
\eeq
with $\ep_0\sim \fr{M'}{M}\ep_e$. Since
$\ep_0$  can be much larger than $\ep_e$ (provided that $\fr{M'}{M}\gg 1$),
we retain $\ep_0$ entries in the $M_N$.
The unitary matrix $U_l$ in (\ref{mD-MN-0}) arises from the choice of basis in which the charged lepton mass matrix is diagonal. It is important to note that
the $2-3$ rotation of $U_l$ [the last multiplier matrix of $U_l$ in Eq. (\ref{mE-diag})] does not alter the structure of $m_D^{(0)}$ and can,
therefore, be absorbed into the redefinition of $\lambda $ and $a_0$. However, the $1-2$ and $1-3$ rotations of $U_l$
do affect the structure of the Dirac neutrino matrix $m_D$.  Using (\ref{mass-e-mu-tau}) and (\ref{mE-diag}), for these rotation angles we have
$$
\te^{~\!\!l}_{13}\simeq \left | \fr{\al''\ep_e}{\ga }\right |=|\al''|(0.1-1)\cdot 10^{-3},
$$
\beq
\te^{~\!\!l}_{12}\simeq \left | \fr{\al'\ep_e}{\bt }\right |\simeq \fr{1}{\lam_{\mu }}|\al'\ep_e|
=|\al'|(0.2-2)\cdot 10^{-2}.
\la{te-ls}
\eeq
Assuming that the values of the couplings $\al', \al''$ can be within the range $1/5-5$,
the largest and most relevant contribution is expected to be due to the $\te^{~\!\!l}_{12}$ (it can be as large as $\sim 0.1$).
Therefore, for $U_l$ we will consider the approximation
\beq
U_l\simeq \!\left(
      \begin{array}{ccc}
        c_e & s_ee^{i\eta }& 0\\
        -s_ee^{-i\eta } & c_e & 0 \\
       0 & 0 & 1 \\
      \end{array}
    \right) ,~~{\rm where}~ \cos \te_e\equiv c_e, ~\sin \te_e\equiv s_e,
\la{Ul-aprox}
\eeq
where we have introduced the angle $\te_e \sim \te^{~\!\!l}_{12}$ and $\eta $ is some phase.

Since both $\ep_0$ and $\te_e$ emerge by $Z_2^T$ symmetry breaking and we assume that the latter symmetry is mildly violated (i.e., the value
of $e^{-\pi y}\ll 1$), we will require that $\ep_0, \te_e\ll 1$. As we will see, this requirement and forms of $m_D, M_N$ and $U_l$ given in
 (\ref{mD-MN-0}) and (\ref{Ul-aprox})
allow only for an inverted ordering (IO) scenario for the light neutrinos.
Applying the seesaw formula $M_{\nu}=-m_DM_N^{-1}m_D^T$, we will get
\beq
M_{\nu}=\!\left(
      \begin{array}{ccc}
        b & a (\ep +\ep_1)& \ep \\
        a(\ep +\ep_1) & a^2(1+\fr{\ep_1^{2}}{b-\ep^2}) & a \\
       \ep & a & 1 \\
      \end{array}
    \right)\!\!\bar m ,
\la{Z2T-nu-matrix}
\eeq

where
$$
\bar m=-\fr{\lam^2v^2}{M(n-\ep_0^2)}~,~~a=a_0c_e-b_0\ep_0s_ee^{-i\eta }~,~\ep=b_0c_e\ep_0+a_0s_ee^{i\eta },
$$
\beq
b=(b_0c_e)^2(n-\ep_0^2)+(b_0c_e\ep_0+a_0s_ee^{i\eta })^2,~~~\ep_1=\fr{\ep^2-b}{a}e^{-i\eta }\tan\te_e .
\la{ab-ab0}
\eeq
%
As noted, within our setup the texture (\ref{Z2T-nu-matrix}) permits the IO  neutrino masses. This can be
demonstrated as follows.
Having chosen the basis in which the charged lepton mass matrix is diagonal, the lepton mixing originates solely from $M_{\nu }$.
Thus, we can write
\beq
 M_{\nu }=PU^*P'M_{\nu}^{\rm Diag}U^{\dag }P,
\la{barMnu}
\eeq
where $U$ represents the lepton mixing matrix in the standard parametrization, and $P, P'$  are the diagonal phase matrices
 [see Eqs. (\ref{Ulept}) and (\ref{Ps}) of Appendix \ref{ApU}]. Utilizing (\ref{Z2T-nu-matrix}) and (\ref{barMnu}), we can express
 $\bar m, a, b, \ep $, and  $\ep_1$ in terms of the entries of $U, P, P'$ [see relations in Eq. (\ref{nu-pars-A}) 
 and the discussion in Appendix \ref{ApU}]. Subsequently, using (\ref{ab-ab0}), we can compute the original parameters.
 Specifically, for the parameter $\te_e$, we find the following results for IO and normal ordering (NO)
 neutrino scenarios:
$$
{\rm For~ IO}\!:~~~\tan \te_e =\fr{s_{13}}{c_{13}s_{23}}\simeq 0.2 ,
$$
\beq
\hs{-0.1cm}{\rm For~ NO}\!:~~~\tan \te_e =\fr{c_{12}c_{13}}{|s_{12}c_{23}+e^{i\de}c_{12}s_{23}s_{13}|}\simeq 1.68-2.56 .
\la{tee-values}
\eeq
For the numerical estimates of (\ref{tee-values}), we used the central values of $\te_{ij}$  \cite{Capozzi:2021fjo} and varied $\de $ within a range $[0, 2\pi[$. This analysis indicates
that, for the normal ordering, $\te_e$ is$\sim 1$. This would correspond to the strong breaking of $Z_2^T$ - a scenario we are
 not interested in exploring.

Therefore, with a focus on the IO case, we present the parameter choices that produce the desired fit.

\begin{figure}[!t]
\begin{center}
\includegraphics[width=0.5\columnwidth]{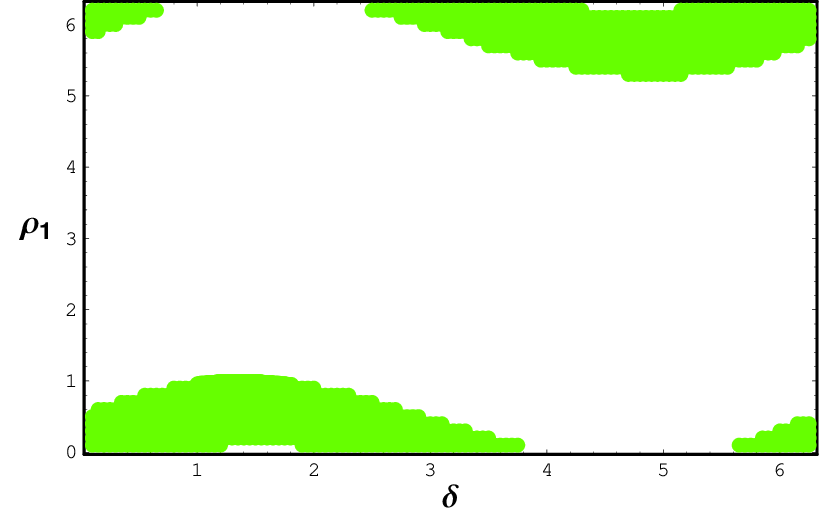}
\vs{-0.3cm}
\caption{Values of $\de $ and $\rho_1$  (green shade) for IO neutrinos, giving $|\fr{\ep_0}{\sqrt{n}}|\leq 0.2$ [see, e.g., Eq. (\ref{orig-pars})].
 We used the best fit values  of the oscillation parameters \cite{Capozzi:2021fjo}.
}
\vs{-0.3cm}
\label{fig1}
\end{center}
\end{figure}

\vs{0.3cm}
{\bf Fit for the IO Case}
\vs{0.3cm}

With the selection
$$
 m=0.028804 ~{\rm eV} ,
$$
\beq
\{a, b, \ep , \ep_1 \}=
\{0.83518, 1.621, -0.002092 + 0.16686 i, -0.017033 + 0.395 i\},
\la{inp-IO}
\eeq
from (\ref{Z2T-nu-matrix})
for the light neutrino masses and mixing angles we obtain
\beq
\{m_1, m_2, m_3 \}=\{0.04917, ~0.04992,~ 0\}{\rm eV},
\la{nu-masses}
\eeq
\beq
\{\sin^2\te_{12},  \sin^2\te_{23}, \sin^2\te_{13}\}=\{0.3035,~ 0.57,~ 0.02235\}.
\la{nu-mixings}
\eeq
{}From (\ref{nu-masses}) we get
\beq
\De m_{\rm sol}^2=m_2^2-m_1^2=7.39\tm 10^{-5}{\rm eV}^2,~~
\De m_{\rm atm}^2=m_2^2= 2.492\tm 10^{-3}{\rm eV}^2~.
\la{nu-mass2dif} 
\eeq
Results of (\ref{nu-mixings}) and (\ref{nu-mass2dif})
correspond to the best fit values  of the inverted ordering neutrino scenario \cite{Capozzi:2021fjo}.
Moreover, for the phases we get
\beq
\{\de, ~\rho_1 \}\simeq \{1.523\pi  , 1.821\pi  \}.
\la{phases}
\eeq
(Since $m_3=0$, the phase $\rho_2$ is unphysical.)
For the parameter choice given above, the neutrinoless double $\bt $-decay parameter
$m_{\bt \bt}=|\sum U_{ei}^2m_iP^{\!\!~'\!*}_i|$  is $m_{\bt \bt}=0.0467$~eV.

Once we have found the parameters giving the good fit, we can go back and see
what are the values of the corresponding original parameters. (See the discussion in Appendix \ref{ApU}.)
Our interest is in those that correspond to the  $Z_2^T$ breaking. Using (\ref{inp-IO}) in Eq. (\ref{orig-pars}),  we get
\beq
\tan\te_e\simeq 0.2~~,~~|\ep_0|\simeq 0.0069\tm \sqrt{|n|}.
\la{tee-ep0}
\eeq
As we see, for $|n|\sim 1$ both parameters are reasonably small, justifying our assumptions pointed out above.
While $\te_e$ is uniquely determined by (\ref{tee-values}),
$|\ep_0|$ depends on  $\de $ and $\rho_1$, fixing their correlation. Figure \ref{fig1}  displays the $(\de, \rho_1)$ values
that yield $\left |\fr{\ep_0}{\sqrt{|n|}}\right | \leq 0.2$.
With $\ep_0\sim \fr{M'}{M}\ep_e$, and $\ep_e\sim 10^{-5}$, the value of $\ep_0$ given in (\ref{tee-ep0}), with $n=1/2$, can be obtained for
the mass ratio $\fr{M'}{M}\sim 5\cdot 10^2$.
Finally, for $|\lam |\!=\!|n|\!=\!1/2$ the selection of  (\ref{inp-IO}), for the heavy RHN masses, gives
\beq
\{M_1, M_2\}=\{ 2.63,~ 5.26\}\tm 10^{14}~{\rm GeV} .
\la{M-N-iinf}
\eeq

\section{Models Near $\tau =i$ Fixed point: ``Intermediate" $Z_4^{S}$  Symmetry}
\label{sec-Fix-i}

Since we are considering even-weighted modular forms $f^{(2n)}(\tau )$, as noted in \cite{Feruglio:2021dte}, 
at $\tau =i$ fixed point  the $S$ transformation,
determined by (\ref{Dk-transf}),
$f^{(2n)}\to (-1)^n\rho(S)f^{(2n)}=\pm \rho (S)f^{(2n)}$ acts as $Z_2^S$ symmetry. However, the fields $\psi $ can have odd weight
and, therefore, in the field space the $S$ transformation can act as  the $Z_4^{S}$ symmetry \cite{Feruglio:2021dte,Feruglio:2023mii}.

As shown in Appendix \ref{ap-Rep-i},
at the fixed point $\tau =i$   there are only two independent (and real) modular doublets
 with weights $(4n+2)$ and $4n$, which possess the following structures:
\beq
{\rm at}~\tau =i:~~~~~ D^{(4n+2)}=\l \!\!\begin{array}{c}

           1 \\
          \fr{1}{\sqrt{3}}
        \end{array}\!\!\r ,~~~~~D^{(4n)}=\l \!\!\begin{array}{c}

           1 \\
          -\sqrt{3}
        \end{array}\!\!\r .
\la{Ds-at-i}
\eeq
It is easy to check that $S$ transformation leaves them invariant, 
\beq
D^{(4n+2)}(i)=i^2\rho_{\bf [2]}(S)D^{(4n+2)}(i),~~~~~D^{(4n)}(i)=\rho_{\bf [2]}(S)D^{(4n)}(i),
\la{Z2S-Dinv}
\eeq
where $\rho_{\bf [2]}(S)$ is given in (\ref{rho2-TS}).

The $k$ weight  fermion state $\psi^{(k)} =(\psi_1, \psi_2)^{T}$ in the doublet representation 
of $S_3$, under the  $Z_4^{S}$ symmetry, transforms as

\beq
{\rm at}~\tau=i:~~~~\psi^{(k)} =\l \!\!\begin{array}{c}
           \psi_1 \\
          \psi_2
        \end{array}\!\!\r \to -\fr{i^k}{2}\!\l \!\!\begin{array}{c}
           \psi_1+\sqrt{3}\psi_2 \\
          \sqrt{3}\psi_1- \psi_2
        \end{array}\!\!\r .
\la{psi2-Z4}
\eeq
Besides these, at $\tau =i$, among the modular forms $\tl E_2$ and $E_{4,6}$ only $E_4$ has nonzero value (\ref{tau-i-E246}).
Also, as discussed in Appendix \ref{ap-Rep-i},
 any singlet modular form of weight $(4n+2)$ and any pseudosinglet  modular form  of weight $4n$ vanish at $\tau=i$,
\beq
{\rm at}~\tau =i:~~~~~ {\bf 1}^{(4n+2)}=0,~~~~~~({\bf 1'})^{(4n)}=0.
\la{1-1pr-i}
\eeq

A few additional useful properties observed at $\tau = i$ arise from the structure presented in (\ref{Ds-at-i}) and the multiplication rule in (\ref{2by2}). Specifically, for the doublet field $\psi = (\psi_1, \psi_2)$ and the modular doublets, we have:
\beq
{\rm at}~\tau =i:~~~~~~(\psi D^{(4n)})_{1}=\sqrt{3}(\psi D^{(4n+2)})_{1'},~~~
(\psi D^{(4n)})_{1'}=-\sqrt{3}(\psi D^{(4n+2)})_{1} .
\la{prop-i}
\eeq
These properties will be utilized in the upcoming model building.
Below we present two models, referred to as Model A and Model B.

\subsection{Model A}
\label{sec-modelA}

For this model, the transformation properties of the SM leptons and the Higgs field  under $S_3$ are the same as those given in (\ref{S3-reps}).
As far as the RHN states are concerned, we still introduce two of them, but now embed them in the $S_3$ doublet,
\beq
S_3:~~~~~~~N=\left(\!\!
      \begin{array}{c}
        N_1 \\
        N_2 \\
      \end{array}\!\!\right)\sim {\bf 2}.
\la{N-S3}
\eeq
In this case, the weights are chosen as follows:
\beq
k_L=k_{e^c_i}=k_N=-1,~~~k_{l_3}=1,~~~k_{\varphi }=0.
\la{weights-i}
\eeq
With these, the couplings relevant for the charged lepton masses will be
$$
-{\cal L}_E= \ga \tl{\varphi }l_3e^c_3-
\tl{\varphi }\left \{(LD^{(4n+2)})_{1'}{\bf 1}^{(-4n)}+(LD^{(4n)})_{1}({\bf 1'})^{(2-4n)}\right \}\!(\bt_0e^c_2+\bt_0'e^c_3)
$$
\beq
+\tl{\varphi }\left \{(LD^{(4n+2)})_{1}({\bf 1'})^{(-4n)}+(LD^{(4n)})_{1'}{\bf 1}^{(2-4n)}\right \}\!(\al_0e^c_1+\al_0'e^c_2+\al_0''e^c_3).
\la{ll-yuk-i}
\eeq
Without loss of any generality
one can choose the basis in which $e^c_3$ couples only with $l_3$. In the second and third terms of Eq. (\ref{ll-yuk-i}) we have given
only relevant invariants.

The couplings responsible for the neutrino sector are
$$
-{\cal L}_{\nu }=\varphi l_3\!\left \{(ND^{(4n)})_{1'}{\bf 1}^{(-4n)}+(ND^{(4n+2)})_{1}{\bf 1'}^{(-4n-2)} \right \}+
$$
$$
\varphi \left \{(LD^{(4n+2)}N)_{1}{\bf 1}^{(-4n)}+(LD^{(4n)}N)_{1'}({\bf 1'})^{(2-4n)}+(LN)_{1'}({\bf 1'})^{(2)}\right \}+
$$
$$
M\left \{ (ND^{(4n+2)}N)_{1}{\bf 1}^{(-4n)}+(ND^{(4n)}N)_{1'}({\bf 1'})^{(2-4n)} \right \}+
$$
\beq
M'\left \{ (NN)_{1}{\bf 1}^{(2)}+ (ND^{(4n)}N)_{1}{\bf 1}^{(2-4n)}+(ND^{(4n+2)}N)_{1'}({\bf 1'})^{(-4n)} \right \}.
\la{lN-yuk-i}
\eeq
In the Dirac-type couplings of (\ref{lN-yuk-i}) we have not included  terms that give  $Z_4^S$ symmetry breaking effects.
The coupling $\ga$ in (\ref{ll-yuk-i}) is responsible for the mass of the $\tau$ lepton, while $\bt_0$ is responsible for the generation of the muon mass. These occur even in the limit of unbroken $Z_4^S$ symmetry.  The mass of the electron will arise from the mild breaking of the $Z_4^S$ symmetry, achieved through a slight deviation from the point $\tau = i \to i + \ep_e$. Consequently, the structures in (\ref{1-1pr-i}), (\ref{Ds-at-i}) and the relations in (\ref{prop-i}) will  also shift
by $\sim \ep_e$.
In Appendix \ref{app-kin} we give the potential, with properly selected parameters, giving the needed value of $\lan \tau \ran$ near the fixed point
$\tau =i$.

  Taking into account all these and relations of (\ref{prop-i}), the Dirac and Majorana mass matrices obtained from (\ref{ll-yuk-i}) and (\ref{lN-yuk-i}) will be:
 \beq
M_E^{(0)}=\!\left(
      \begin{array}{ccc}
        \al \ep_e &-\fr{\bt}{\sqrt{3}}+ \al'\ep_e   & -\fr{\bt'}{\sqrt{3}}+\al'' \ep_e \\
      \al \fr{\ep_e}{\sqrt{3}}  & \bt+ \al' \fr{\ep_e}{\sqrt{3}}  & \bt'+ \al'' \fr{\ep_e}{\sqrt{3}}  \\
       0 & 0 & \fr{2}{\sqrt{3}}\ga\\
      \end{array}
    \right)\!\!\fr{\sqrt{3}v}{2} +{\cal O}(\ep_e^2),
\la{ME-at-i}
\eeq
\beq
{m_D^{(0)}=\!\!\left(
\begin{array}{ll}
 -a & \frac{a}{\sqrt{3}}+b \\
 \frac{a}{\sqrt{3}}-b  & a \\
 \sqrt{3} & 1
\end{array}
\right) \!\!\lam v,~~\text{   }M_N=\!\!\left(\!
\begin{array}{cc}
 -1+n \epsilon_0  & \frac{1}{\sqrt{3}}+\epsilon_0   \\
 \frac{1}{\sqrt{3}}+\epsilon_0  & 1+n \epsilon_0
\end{array}
\! \right) \!\!M.}\\
\la{mDMN-at-i}
\eeq
[As noted, in the $m_D^{(0)}$ the entries$\sim \ep_e$  (i.e., the $Z_4^S$ breaking effects) are neglected. The latter parameter
will come out to be $\ep_e\sim \lam_e$.]
In $M_N$, since the parameter $\ep_0\sim \fr{M'}{M}\ep_e$  can be much larger than $\ep_e$ (provided that $\fr{M'}{M}\gg 1$),
we retain the corresponding entries.
Furthermore,  making the $l_1-l_2$ rotation, i.e. going to the basis $l\to U_0^Tl$, where
 \beq
 U_0=\!\left(
      \begin{array}{ccc}
        \fr{\sqrt{3}}{2} & \fr{1}{2} & 0 \\
       \!\! -\fr{1}{2} &\fr{\sqrt{3}}{2} & 0  \\
       0 & 0& 1 \\
      \end{array}
    \right),
 \la{UZT}
 \eeq
 the Dirac-type matrices will become
 \beq
M_E=U_0M_E^{(0)}=\!\left(
      \begin{array}{ccc}
        \al \ep_e & \al'\ep_e   & \al'' \ep_e \\
      0  & \bt   & \bt'   \\
       0 & 0 & \ga  \\
      \end{array}
    \right)\!\!v  ,~~~
m_D^{(1)}=U_0m_D^{(0)}.
\la{UZME-MD}
\eeq
Since the members $\nu $ and $e^{-}$  of the $l$
doublets get rotated by the same  unitary matrix $U_0$, the latter does not contribute to the lepton mixing matrix.
The form of $M_E$ coincides with the one given in Eq. (\ref{ME}) obtained and analyzed in the model presented in Sec. \ref{ME-inf}.
Therefore, for the details of  $M_E$'s study we refer to this section. As before, from the diagonalization  of $M_E$ we will take the form
of (\ref{Ul-aprox}) and make further transformation of the $l$ states by this matrix. Therefore, in this basis the neutrino's Dirac matrix becomes
\beq
m_D=U_lm_D^{(1)}=U_lU_0m_D^{(0)}.
\la{mD-fin}
\eeq
Applying the seesaw formula $M_{\nu }=-m_DM_N^{-1}m_D^T$ and using forms of (\ref{mDMN-at-i}), (\ref{UZT}), (\ref{Ul-aprox}) and  (\ref{mD-fin}),
for the neutrino mass matrix we obtain
\beq
M_{\nu}=\!{U_l\!\!\left(
\begin{array}{lll}
 \varepsilon _2 & \bar a & 1 \\
\bar a& \bar a^2 \varepsilon _1 &\bar a \varepsilon _1 \\
 1 & \bar a \varepsilon _1 & \varepsilon _1
\end{array}
\right)\!\!U_l^T}\bar m,\\
\pmb{}\\
\la{Mnu-i-A}
\eeq
where
$$
\bar a=\frac{a}{\sqrt{3}}-\frac{b}{2},~~\varepsilon_1=\frac{6(\sqrt{3}-2 n)}{(2a+\sqrt{3}b)(4+\sqrt{3}\ep_0 )}\ep_0 ,~~
\varepsilon_2=-\frac{(2a+\sqrt{3}b) \left(\sqrt{3}+2 n\right) }{2(4+\sqrt{3} \ep_0 )}\ep_0 ,~~
$$
\beq
\bar m=-\fr{(\lam v)^2}{M}\fr{(2a+\sqrt{3}b)(4+\sqrt{3}\ep_0)}{4+2\sqrt{3}\ep_0+3(1-n^2)\ep_0^2}~.
\la{not-i}
\eeq
In a limit $\ep_0\to 0, \te_e\to 0$, Eq. (\ref{Mnu-i-A}) gives bilarge neutrino mixing with the IO neutrinos, but with $\De m_{\rm sol}^2=0, \te_{13}=0$,
and $\te_{12}=\fr{\pi}{4}$.
Therefore, the $Z_4^S$ breaking effects are important for the realistic model.\footnote{In \cite{Shafi:2000su}, a similar texture
was obtained with a ${\cal U}(1)$ flavor symmetry, in which symmetry breaking effects played an important role in realistic neutrino oscillations.}
With the selection
 \beq
 \{\bar a, \varepsilon _1, \varepsilon _2, \te_e, \eta \}= \{ 0.85266, 0.11718, -0.1824, 0.19775, \pi  \}, ~~~|\bar m|=0.037307~{\rm eV} ,
 \la{select1}
 \eeq
the texture (\ref{Mnu-i-A}) gives the IO neutrino masses, which coincides with those given in Eq. (\ref{nu-masses})
with the best-fit values of the oscillation parameters of Eqs. (\ref{nu-mixings}) and (\ref{nu-mass2dif}).
The neutrinoless double $\bt $-decay parameter is $m_{\bt \bt }=0.0187$~eV.

The choice (\ref{select1}), with the relations of Eq. (\ref{not-i}) for the original parameters gives the natural selection:
\beq
\{a, b, n, \epsilon_0 \}=\{1.2472e^{-0.4864i}, -0.8e^{i}, 1.187e^{0.71586i}, 0.14484e^{0.63142i} \} .
\la{select2}
 \eeq
With these, for $|\lam |\simeq 0.31$ we obtain $|M|\simeq 2\cdot 10^{14}$~GeV, which for the heavy RHN masses gives
\beq
\{M_1, M_2\}=\{ 2.29,~ 2.62\}\tm 10^{14}~{\rm GeV} .
\la{M-N-iA}
\eeq

\subsection{Model B}
\label{sec-modelB}

Within this model, two RHNs, $N$ and $N'$, are still introduced
The transformation properties of the states under $S_3$ are the same as those given in (\ref{S3-reps}) and (\ref{N-transf}),
while  the weights are chosen as follows:
\beq
k_L=-2,~~~k_{l_3}=k_{e^c_i}=k_N=k_{N'}=k_{\varphi }=0.
\la{weights-i-B}
\eeq
With these, the couplings relevant for the charged lepton masses will be
$$
-{\cal L}_E=\ga \tl{\varphi }l_3e^c_3-
\tl{\varphi }\left \{(LD^{(4n+2)})_{1'}{\bf 1}^{(-4n)}+(LD^{(4n)})_{1}({\bf 1'})^{(2-4n)}\right \}\!(\bt_0e^c_2+\bt_0'e^c_3)
$$
\beq
+\tl{\varphi }\left \{(LD^{(4n+2)})_{1}({\bf 1'})^{(-4n)}+(LD^{(4n)})_{1'}{\bf 1}^{(2-4n)}\right \}\!(\al_0e^c_1+\al_0'e^c_2+\al_0''e^c_3)~,
\la{ll-yuk-i-B}
\eeq
while the couplings in the neutrino sector are
$$
-{\cal L}_{\nu }= \varphi l_3\!\left \{ N' {\bf 1}^{(0)}+N ({\bf 1'})^{(0)}\right \}+
$$
$$
\varphi \left \{(LD^{(4n+2)})_{1}{\bf 1}^{(-4n)}+(LD^{(4n)})_{1'}({\bf 1'})^{(2-4n)}+(LD^{(4n)})_{1}({\bf 1})^{(2-4n)}
+(LD^{(4n+2)})_{1'}({\bf 1'})^{(-4n)}\right \}\!N+
$$
$$
\varphi \left \{(LD^{(4n+2)})_{1'}{\bf 1}^{(-4n)}+(LD^{(4n)})_{1}({\bf 1'})^{(2-4n)}+(LD^{(4n)})_{1'}({\bf 1})^{(2-4n)}
+(LD^{(4n+2)})_{1}({\bf 1'})^{(-4n)}\right \}\!N' +
$$
\beq
\fr{1}{2}M\left \{N^2{\bf 1}^{(0)}+(N')^2{\bf 1}^{(0)}+2NN'({\bf 1'})^{(0)} \right \}.
\la{lN-yuk-i-B}
\eeq
Using in (\ref{ll-yuk-i-B}) and (\ref{lN-yuk-i-B}) the structures of (\ref{1-1pr-i}) and (\ref{Ds-at-i}), and the relations  (\ref{prop-i})
(shifted by $\sim \ep_e$ due to $Z_4^S$ breaking effects),
and at the same time disregarding the effects of $Z_4^S$ breaking in the neutrino Dirac-type couplings
(as in  Model A, discussed in Sec. \ref{sec-modelA}),
the Dirac and Majorana mass matrices  will be
\beq
M_E^{(0)}=\!\left(
      \begin{array}{ccc}
        \al \ep_e &-\fr{\bt}{\sqrt{3}}+ \al'\ep_e   & -\fr{\bt'}{\sqrt{3}}+\al'' \ep_e \\
      \al \fr{\ep_e}{\sqrt{3}}  & \bt+ \al' \fr{\ep_e}{\sqrt{3}}  & \bt'+ \al'' \fr{\ep_e}{\sqrt{3}}  \\
       0 & 0 & \fr{2}{\sqrt{3}}\ga\\
      \end{array}
    \right)\!\!\fr{\sqrt{3}v}{2} +{\cal O}(\ep^2),
\la{ME-at-i-B}
\eeq

\beq
m_D^{(0)}=\!\left(
      \begin{array}{cc}
        \sqrt{3}b & -a \\
        b & \sqrt{3}a  \\
       0 & 2 \\
      \end{array}
    \right)\!\! \fr{\lam v}{2} ,~~~
M_N=\!\left(
      \begin{array}{cc}
        1 & -\ep_0 \\
        -\ep_0 & n  \\
      \end{array}
    \right)\!\! M .
\la{mDMN-at-i-B}
\eeq
(As noted, in $m_D$ are neglected  entries$\sim \ep_e$. The latter parameter will come out to be $\ep_e\sim \lam_e$.)
Furthermore, as we have done for Model A (of Sec. \ref{sec-modelA}), making the $l_1-l_2$ rotation, i.e. going to the basis $l\to U_0^Tl$
[where $U_0$ is given in Eq. (\ref{UZT})], the Dirac-type matrices will become
 \beq
M_E=U_0M_E^{(0)}=\!\left(
      \begin{array}{ccc}
        \al \ep_e & \al'\ep_e   & \al'' \ep_e \\
      0  & \bt   & \bt'   \\
       0 & 0 & \ga  \\
      \end{array}
    \right)\!\!v  ,~~~
m_D=U_0m_D^{(0)}=\!\left(
      \begin{array}{cc}
        b & 0 \\
        0 & a  \\
       0 & 1 \\
      \end{array}
    \right)\!\! \lam v .
\la{UZME-MD-B}
\eeq
Since $U_0$ transforms the entire lepton doublet states $l_{1,2,3}$, the lepton mixing matrix is  determined by the diagonalization
of the matrices $M_E$ and $M_{\nu }$.
As we see, the matrices in Eq. (\ref{UZME-MD-B}) and $M_N$ are the same  as those in  (\ref{ME}) and (\ref{mD-MN-0})
(obtained near the $\tau =i\infty $ fixed point).
Therefore, the results (e.g., charged lepton masses and the neutrino fit) and conclusions will be essentially
 the same as those previously discussed in Sec. \ref{sec-Fix-inf}. Thus, the IO neutrino scenario remains the preferred choice within this model as well.

\section{Model Near $\tau =\om $ Fixed point: ``Intermediate" $Z_3^{ST}$  Symmetry}
\label{sec-Fix-om}

At $\tau \!=\!\om \!=\!-\fr{1}{2}\!+\!i\fr{\sq{3}}{2}$ fixed point we have residual $Z_3^{ST}$ symmetry
\cite{Feruglio:2023mii,Novichkov:2018ovf,Ding:2019gof,Okada:2020ukr}
realized by the $ST$ transformation.
Under this transformation, at this point, the modular form $f^{(k)}$
transforms as  
$f^{(k)}\xrightarrow{ST}\om^{2k}\rho(ST)f^{(k)}$. Thus, the elements of $Z_3^{ST}$ are
$g_i(Z_3^{ST})=\{ 1, ST, (ST)^2 \}$.

As shown in Appendix \ref{ap-Rep-om},
at  $\tau =\om $   the modular doublets
 with weights $(6n+p)$ (where $p=0, 2, 4$) possess the following structures:
\beq
{\rm at}~\tau =\om :~~~~~ D^{(6n)}=\l \!\!\begin{array}{c}

           0 \\
          0
        \end{array}\!\!\r ,~~~~D^{(6n+2)}=\l \!\!\begin{array}{c}

           1 \\
          -i
        \end{array}\!\!\r ,~~~~D^{(6n+4)}=\l \!\!\begin{array}{c}

           1 \\
          i
        \end{array}\!\!\r .
\la{Ds-at-om}
\eeq
It is easy to check that $ST$ transformation leaves them invariant, 
\beq
D^{(6n+p)}(\om )=\om^{2p}\rho_{\bf [2]}(S)\rho_{\bf [2]}(T)D^{(6n+p)}(\om ),~~~~p=0, 2, 4,
\la{Z3-Dinv}
\eeq
where $\rho_{\bf [2]}(S)$ and $\rho_{\bf [2]}(T)$ are given in (\ref{rho2-TS}).

Besides these, at $\tau =\om $, among the modular forms $\tl E_2$ and $E_{4,6}$ only $E_6$ has nonzero value (\ref{tau-om-E246}).
Also, as discussed in Appendix \ref{ap-Rep-om}, at $\tau=\om $
 the singlet and  pseudosinglet modular forms of different weights are 

\begin{eqnarray}
             \nonumber {\rm at}~\tau =\om :~~ & {\bf 1}^{(6n)}\neq 0, & ~{\bf 1}^{(6n+2)}={\bf 1}^{(6n+4)}=0, \\
               & ~({\bf 1'})^{(6n)}\neq 0,& ~ ({\bf 1'})^{(6n+2)}=~({\bf 1'})^{(6n+4)}=0 . 
\la{1-1pr-om}
\end{eqnarray}

Additional properties useful for model building are provided by the following relations, which involve the doublet field $\psi = (\psi_1, \psi_2)$ and the modular doublets:
\beq
{\rm at}~\tau =\om:~~~~~~(\psi D^{(6n+2)})_{1}=i(\psi D^{(6n+2)})_{1'},~~~
(\psi D^{(6n+4)})_{1'}=i(\psi D^{(6n+4)})_{1} .
\la{prop-om}
\eeq
The following subsection details Model C in the neighborhood of the $\tau = \om $ fixed point.

\subsection{Model C}
\label{sec-modelC}

For this model, the transformation properties of the SM leptons and the Higgs field  under $S_3$ are the same as those given in (\ref{S3-reps}).
Two RHN states are  embedded in the $S_3$ doublet, as given in (\ref{N-S3}), and
 the weights are chosen as follows:
\beq
k_L=k_{e^c_i}=-1,~~~k_{l_3}=k_N=1,~~~k_{\varphi }=0.
\la{weights-om}
\eeq
With these, the couplings relevant for the charged lepton masses will be
$$
-{\cal L}_E= \ga \tl{\varphi }l_3e^c_3+
\tl{\varphi }\left \{(LD^{(6n+2)})_{1'}{\bf 1}^{(-6n)}+(LD^{(6n+2)})_{1}({\bf 1'})^{(-6n)}\right \}\!(\bt_0e^c_2+\bt_0'e^c_3)
$$
\beq
+\tl{\varphi }\left \{(LD^{(6n+4)})_{1}({\bf 1'})^{(-6n-2)}+(LD^{(6n+4)})_{1'}{\bf 1}^{(-6n-2)}\right \}\!(\al_0e^c_1+\al_0'e^c_2+\al_0''e^c_3).
\la{ll-yuk-om}
\eeq
Invariants 
$(LD^{(6n)})_{1'}{\bf 1}^{(2-6n)}$ and $(LD^{(6n)})_{1}({\bf 1'})^{(2-6n)} $
 are not included, as their contributions lead to negligible couplings of order $\epsilon_e^2$.

The couplings relevant for the neutrino sector are
$$
-{\cal L}_{\nu }=\varphi l_3\!\left \{(ND^{(6n-2)})_{1'}{\bf 1}^{(-6n)}+(ND^{(6n-2)})_{1}{\bf 1'}^{(-6n)} \right \}+
\varphi \left \{(LN)_{1}{\bf 1}^{(0)}+(LN)_{1'}({\bf 1'})^{(0)}\right \}+
$$
$$
M\left \{ (ND^{(6n-2)}N)_{1}{\bf 1}^{(-6n)}+(ND^{(6n-2)}N)_{1'}({\bf 1'})^{(-6n)} \right \}+
$$
\beq
M'\left \{ (NN)_{1}{\bf 1}^{(-2)}+ (ND^{(6n+2)}N)_{1}{\bf 1}^{(-6n-4)}+(ND^{(6n+2)}N)_{1'}({\bf 1'})^{(-6n-4)} \right \}.
\la{lN-yuk-om}
\eeq
In Eq. (\ref{lN-yuk-om}), we have not included $Z_3^{ST}$ symmetry breaking terms within the Dirac-type 
couplings.
In addition, the invariants
$(ND^{(6n)}N)_{1}{\bf 1}^{(-6n-2)}$ and $(ND^{(6n)}N)_{1'}({\bf 1'})^{(-6n-2)}$
have been omitted, as they provide corrections of order$\sim \epsilon^2$.

 Similar to the models presented in previous sections, the electron mass in this case arises 
 from the mild breaking of the $Z_3^{ST}$ symmetry, induced by a tiny deviation from the
  fixed point $\tau = \omega \to \omega + \epsilon_e$.
 Consequently, the structures in (\ref{1-1pr-om}) and (\ref{Ds-at-om}), as well as the relations in (\ref{prop-om}), will also be shifted by $\sim \epsilon_e$.
 How the needed value of $\langle \tau \rangle$ is obtained near the fixed point $\tau = \omega$ is
 addressed in Appendix \ref{app-kin}.
 The couplings $\gamma$ and $\beta_0$ in (\ref{ll-yuk-om}) are responsible for the masses of the $\tau$ lepton and the muon, respectively.

  Taking into account all the above, the Dirac and Majorana mass matrices obtained from (\ref{ll-yuk-om}) and (\ref{lN-yuk-om}) will be:
 \beq
M_E^{(0)}=\!\left(
      \begin{array}{ccc}
        \al \ep_e &\bt +\al'\ep_e   & \bt'+\al'' \ep_e \\
      i\al \ep_e & -i\bt+ i\al' \ep_e & -i\bt'+i \al'' \ep_e  \\
       0 & 0 & \sqrt{2}\ga\\
      \end{array}
    \right)\!\!\fr{v}{\sqrt{2}} +{\cal O}(\ep_e^2),
\la{ME-at-om}
\eeq
\beq
{m_D^{(0)}=\!\!\left(
\begin{array}{ll}
 a_0 & b_0 \\
 \!\!\!\!-b_0  & a_0 \\
 1& i
\end{array}
\right) \!\!\lam v,~~M_N=\!\!\left(\!
\begin{array}{cc}
 -1+\epsilon_0  & i+n_0\epsilon_0   \\
 i+n_0\epsilon_0  & 1+\epsilon_0
\end{array}
\! \right) \!\!M.}\\
\la{mDMN-at-om}
\eeq
 The parameter $\ep_e$ will come out to be $\ep_e\sim \lam_e$.
 With $\fr{M'}{M}\gg 1$,
in $M_N$ the parameter $\ep_0\sim \fr{M'}{M}\ep_e$  can be much larger than $\ep_e$.

Now,  going to the basis $l\to U_0^Tl$, with
 \beq
 U_0=\!\left(
      \begin{array}{ccc}
        \fr{1}{\sqrt{2}} & -\fr{i}{\sqrt{2}} & 0 \\
       \!\! \fr{1}{\sqrt{2}}  &\fr{i}{\sqrt{2}}  & 0  \\
       0 & 0& 1 \\
      \end{array}
    \right)
 \la{UZT-om}
 \eeq
 (i.e., performing the $l_1-l_2$ rotation),
 the Dirac-type matrices will become
 \beq
M_E=U_0M_E^{(0)}=\!\left(
      \begin{array}{ccc}
        \al \ep_e & \al'\ep_e   & \al'' \ep_e \\
      0  & \bt   & \bt'   \\
       0 & 0 & \ga  \\
      \end{array}
    \right)\!\!v  ,~~~
m_D^{(1)}=U_0m_D^{(0)}.
\la{UZME-MD-om}
\eeq
The form of $M_E$ coincides with the one given in Eq. (\ref{ME}), obtained and analyzed in the model presented in Sec. \ref{ME-inf}.
Thus, for the details we refer to this section.  As before, from the diagonalization  of $M_E$ we will take form
of (\ref{Ul-aprox}) and perform further transformation of the $l$ states by this matrix. Thus, in this basis the neutrino's Dirac matrix becomes
\beq
m_D=U_lm_D^{(1)}=U_lU_0m_D^{(0)}.
\la{mD-fin-om}
\eeq

Using  the  forms of (\ref{mDMN-at-om}), (\ref{UZT-om}), (\ref{Ul-aprox}), and (\ref{mD-fin-om})
in the seesaw formula $M_{\nu }=-m_DM_N^{-1}m_D^T$,
for the neutrino mass matrix we obtain
\beq
M_{\nu}=\!\left(
      \begin{array}{ccc}
        n\ep & a (\ep +\ep_1)& \ep \\
        a(\ep +\ep_1) & a^2(1+\fr{\ep_1^{2}}{n\ep -\ep^2}) & a \\
       \ep & a & 1 \\
      \end{array}
    \right)\!\!\bar m ,
\la{Mnu-om-C}
\eeq
where the parameters of $M_{\nu}$ are expressed in terms of the original parameters as follows:
$$
a\!=\!c_ea_1-e^{-i\eta }s_e\xi ,~~
n\!=\!\fr{e^{2i\eta}s_e^2a_1^2+c_e^2\xi n1+2e^{i\eta}c_es_ea_1\xi}
{e^{i\eta}s_ea_+c_e\xi},
$$
$$
\ep \!=\!e^{i\eta}s_ea_1+c_e\xi ,~~
\ep_1\!=\!\fr{c_es_e(n_1-\xi)}{s_e\xi-e^{i\eta}c_ea_1},~~
\bar m=\fr{(\lam v)^2}{M} \fr{2(in_0\ep_0-2)}{\ep_0^2(1-n_0^2)-2in_0\ep_0},
$$
\beq
{\rm with}~~~a_1=\fr{1}{\sqrt{2}}(a_0-ib_0),~~~
\xi =\fr{1}{\sqrt{2}}\fr{a_0+ib_0}{2-in_0\ep_0}\ep_0,~~
n_1=\fr{i}{\sqrt{2}}(a_0+ib_0)n_0.
\la{not-om}
\eeq
 The texture (\ref{Mnu-om-C}) matches the form of (\ref{Z2T-nu-matrix}) 
(near the $\tau = i\infty$ fixed point) by replacing $n\epsilon$ with $b$. As a result, 
the relations from Eq. (\ref{tee-values}) still hold, 
favoring the IO neutrino mass hierarchy under the condition that $\theta_e$ remains small.

We present a selection of natural parameter values that provide a consistent fit. In particular, the choice 
$$
\{a_0, b_0 , \te_e, \eta \}=\{2.3394,~ 2.9732e^{-i1.9616},~ 0.19765,~ -0.033185 \},
$$
\beq
\{\lam ,n_0 ,  \epsilon_0 \}=\{0.3,~1.2403e^{i2.345},~0.2006e^{i0.94271} \},~~~M=7.306\tm 10^{14} ~{\rm GeV} ,
\la{select-om}
 \eeq
yields oscillation parameters in Eqs. (\ref{nu-masses})–(\ref{nu-mass2dif}), which are in excellent 
agreement with the best-fit values for the IO scenarios.
Note that a slightly larger selection for the values of $a_0$ and $b_0$ does not pose any problems, 
as the physical Yukawa couplings $a_0\lambda, b_0\lambda \stackrel{<}{_\sim} 1$ remain perturbative.
 The selection in (\ref{select-om}) yields 
 $\{\de, \rho_1 \}=\{6.25, 5.55 \}$, $m_{\bt \bt}=0.0456$~eV and 
 results in heavy RHN masses of $\{M_1, M_2\}\simeq \{ 1.95,~ 14.6\}\tm 10^{14}~{\rm GeV}$.

\vs{0.5cm}

\hs{-0.6cm}In summary, we have investigated the SM extension with the minimal modular flavor 
symmetry $\Ga_2 \simeq S_3$ and built the
lepton sector near the $\tau = i\infty$, $\tau = i$, and 
$\tau \!=\!\om \!=\!-\fr{1}{2}\!+\!i\fr{\sqrt{3}}{2}$ fixed points. 
In our construction, the scalar sector is extended solely by a single complex modulus, $\tau$, which serves as the fundamental variable of the modular forms. These forms transform under $S_3$ representations, allowing us to omit traditional flavon scalars from the theory. The fermion sector is similarly minimal, featuring an extension of only two right-handed neutrinos, $N_{1,2}$.
The residual symmetries at the fixed points greatly reduce the number
of independent invariants (in which nonholomorphic modular forms participate) and also allow us to make some predictions.
In the same spirit,  it would be interesting to 
 examine larger modular groups such as $\Ga_{3,4,5,\dots} \simeq A_4, S_4, A_5,\dots$, and 
to include the quark sector as well. 
While the minimal $S_3$ modular group is highly constrained, larger modular groups can
 possess several small expansion parameters, allowing for a successful explanation of the 
 hierarchies among the three generations. These and related issues remain to be pursued elsewhere \cite{in-prep}.

\subsubsection*{Acknowledgments}
I wish to acknowledge the center for theoretical Underground Physics and
Related Areas (CETUP*) and the Institute for Underground Science at Sanford Underground 
Research Facility (SURF), and the South Dakota Science and Technology
 Authority for hospitality and financial support, as well as for providing a stimulating 
 environment during
 the neutrino workshop.

\appendix

\renewcommand{\theequation}{A.\arabic{equation}}\setcounter{equation}{0}

\section{Modular Forms and $\Ga_2\simeq S_3$ Representations}
\la{app-S3}


For building the modular forms, the Eisenstein series play a central role \cite{kaneko, apostol, zagier}. The modular form
$\tilde{E}_2(\tau )$ with the transformation property (\ref{fk}) (with $k = 2$) is:
\beq
\tl E_2(\tau)=E_2(\tau)-\fr{6i}{\pi (\tau-\bar{\tau})}~,
\la{E2tl}
\eeq
where $E_2(\tau )$ is the lowest Eisenstein series,
\beq
E_2(\tau)=1-24\sum_{n=1}^{\infty }\si_1(n)q^n=1-24(q+3q^2+4q^3+7q^4+6q^5+\cdots )~ ,
\la{E2}
\eeq
where $\sigma_1(n)$ denotes the sum of all divisors of $n$, and $q = e^{2\pi i \tau}$.
While $\tl E_2$ is not holomorphic, the $E_4(\tau)$ and $E_6(\tau)$ Eisenstein series, of weights $4$ and $6$, respectively, are
holomorphic modular forms and are given by
\beq
E_4(\tau)=1+240\sum_{n=1}^{\infty }\si_3(n)q^n ~,~~~~~E_6(\tau)=1-504\sum_{n=1}^{\infty }\si_5(n)q^n ,
\la{E4-E6}
\eeq
where $\sigma_p(n)$ denotes the sum of the $p$th powers of all divisors of $n$ [e.g., $\sigma_3(4)=1+2^3+4^3$, $\sigma_5(6)=1+2^5+3^5+6^5$, etc.].

\vs{0.3cm}
{\bf  Group $\Ga_2\simeq S_3$ and It's Properties}
\vs{0.1cm}

\hs{-0.6cm}The finite group $\Ga_2\simeq S_3$ has six elements $g_i$ ($i=1 \div 6$), which are expressed  
by two generating elements $S$ and $T$,
\beq
g(S_3)=\{ 1,~ T,~ S,~ TS,~ ST,~ STS\},
\la{S3-elements}
\eeq
where $S$ and $T$ satisfy
\beq
T^2=S^2=1, ~~(ST)^3=1 .
\la{ST}
\eeq

The group $S_3$ possesses, in addition to the singlet
representation ${\bf 1}$, a pseudosinglet representation ${\bf 1'}$  and a doublet representation
${\bf 2}$.
According to the multiplication rule for  doublets ${\bf 2}\tm {\bf 2}={\bf 2}+{\bf 1}+{\bf 1'}$, we have
\beq
\l \!\!\begin{array}{c}
  A_1 \\
  A_2
\end{array}\!\!\r_{\!\bf{2}}\!\!\tm \!
 \l \!\!\begin{array}{c}
  B_1 \\
  B_2
\end{array}\!\!\r_{\!\bf{2}}\!=\! \l \!\!\begin{array}{c}
  A_2B_2-A_1B_1 \\
  A_1B_2+A_2B_1
\end{array}\!\!\r_{\!\bf{2}}\!+\! (A_1B_1+A_2B_2)_{\bf 1}\!+\! (A_1B_2-A_2B_1)_{\bf 1'}~.
\la{2by2}
\eeq
In addition to (\ref{2by2}), the multiplication rule
\beq
C_{\!\bf 1'}\tm (A_1, A_2)_{\bf 2}=(-CA_2, CA_1)
\la{1prby2}
\eeq
may be useful for model building. Moreover, the following multiplication rules are straightforward:
\beq
{\bf 1}\tm {\bf 1'}={\bf 1'}~~ {\rm and}~~ {\bf 1'}\tm {\bf 1'}={\bf 1}~.
\la{1by1pr}
\eeq
Using these, one can obtain representations ${\bf 2}, {\bf 1}$, and ${\bf 1'}$ of any even weights by multiplying modular forms
with a given weight and  belonging to the  appropriate representations of the $S_3$.

An $S_3$ singlet modular form of any even weight is given by the general expression (\ref{fk-from-kaEs}). In fact, starting from some
modular form of weight $k$ and applying the weight-raising operator $\hat {\bf \pl }$ and the weight-lowering operator $\hat {\bf \de^*}$
\cite{zagier,Nagatomo}, one can construct  modular forms of weights $k+2$ and $k-2$, respectively.
These operators are defined by \cite{zagier,Nagatomo}
$$
 \hat {\bf \pl }:=\!\fr{1}{2\pi i}\l \fr{\pl }{\pl \tau }+\fr{k}{\tau -\ov{\tau}}\r =\!
 \fr{1}{4\pi i}\l \fr{\pl }{\pl x}-i\fr{\pl }{\pl y}-i\fr{k}{y}\r ,
 $$
\beq
\hat {\bf \de^*}:=\!2\pi i (\tau -\ov{\tau})^2\fr{\pl }{\pl \ov{\tau }}=\!-4\pi i y^2\l \fr{\pl }{\pl x}+i\fr{\pl }{\pl y}\r .
 \la{del-de-ops}
 \eeq
By applying these operators in sequence, one can obtain modular forms of any desired weight. For example, taking the weight-$0$ form
$\ka^2|\tl E_2|^2$, and applying these operators, one obtains modular forms of weights $2$ and  $-2$,
$$
 \hat {\bf \pl }(\ka^2|\tl E_2|^2)=\fr{3}{4\pi^2}\tl E_2+\fr{1}{12}\ka^2(\tl E_2^2-E_4)\tl E_2^* ,
$$
\beq
\hat {\bf \de^*}(\ka^2|\tl E_2|^2)=12\ka^2\tl E_2^*+\fr{4\pi^2}{3}\ka^4(\tl E_2^{*2}-E_4^*)\tl E_2 ,
\la{resulting}
\eeq
where the derivatives of modular forms have been evaluated employing Ramanujan's formulae (see Eq. (3) in \cite{Nagatomo}).
 Both resulting modular forms of Eq. (\ref{resulting}) are contained in (\ref{fk-from-kaEs}).

 \vs{0.2cm}

In the doublet representation, two generating  group elements $\rho_{[2]} (T)$ and  $\rho_{[2]} (S)$ are
\beq
\rho_{[2]} (T)= \left(\!
    \begin{array}{cc}
      1 & 0 \\
      0 & -1  \\
    \end{array}\!\right)~,~~~
  \rho_{[2]} (S)=\fr{1}{2}\left(\!
    \begin{array}{cc}
      -1 & -\sqrt{3} \\
      -\sqrt{3} & 1  \\
    \end{array}\!\right)~,
\la{rho2-TS}
\eeq
and the $S_3$ doublet of weight $k$ - $D^{(k)}(\tau )=\l D_1^{(k)}(\tau ), D_2^{(k)}(\tau )\r $ - has the following transformation properties:
$$
T:~~~D^{(k)}(\tau+1 )=\rho_{[2]} (T)D^{(k)}(\tau )~,
$$
\beq
S: ~~~D^{(k)}(-\fr{1}{\tau } )=\tau^k\rho_{[2]} (S)D^{(k)}(\tau )~ .
\la{Dk-transf}
\eeq
The $S_3$ doublet of weight $k$ - $D^{(k)}(\tau )=\l D_1^{(k)}(\tau ), D_2^{(k)}(\tau )\r $  - can be constructed
 form the $f^{(k)}(\tau )$ modular form of weight $k$ [see (\ref{fk})]
 via the functions $f^{(k)}\l \fr{\tau}{2}\r ,f^{(k)}\l \fr{\tau +1}{2}\r $ and $f^{(k)}\l 2\tau \r$ as follows:
$$
D_1^{(k)}(\tau )=-a_k\left [ f^{(k)}\l \fr{\tau}{2}\r +f^{(k)}\l \fr{\tau +1}{2}\r-2^{k+1}f^{(k)}\l 2\tau \r\right ],
$$
\beq
D_2^{(k)}(\tau )=-a_k\sqrt{3}\left [ f^{(k)}\l \fr{\tau}{2}\r -f^{(k)}\l \fr{\tau +1}{2}\r \right ],
\la{Dk}
\eeq
where $a_k$ are some normalization factors and can be selected by convenience.
Using (\ref{fk}), (\ref{rho2-TS}), and  (\ref{Dk}) one can easily verify that the transformations (\ref{Dk-transf}) are indeed acquired.

Using this result,
the weight-2 doublet -denoted by $Y=(Y_1,Y_2)$ - can be built from $\tl{E}_2$. Using in (\ref{Dk})  $f^{(2)}=\tl{E}_2$ and  $a_2=1/48$, we will have
$$
Y_1(\tau ) =1/8+\!3q +\!3 q^2 +\!12 q^3 +\!3 q^4 +\!18 q^5 +\!12 q^6 +\!24 q^7 +\!3 q^8+\!\cdots ,
$$
\beq
Y_2(\tau )=\sqrt{3}e^{i\pi \tau }\l 1 + 4 q + 6 q^2 + 8 q^3 + 13 q^4 + 12 q^5 + 14 q^6 + 24 q^7 + 18 q^8 +\!\cdots\r .
\la{Y12}
\eeq
Note that in $Y_{1,2}$ the nonholomorphic parts cancel out and, therefore, the doublet $Y$ is holomorphic. The  weight-4 holomorphic doublet
 - $Y^{(4)}=(Y_1^{(4)},Y_2^{(4)})$ - can be built by taking $f^{(4)}=E_4$ and using it in (\ref{Dk}) with $k=4$. As a result (by selecting
 $a_4=-1/480$), we obtain

 $$
Y_1^{(4)}(\tau ) =-1/16+\!9q +\!57 q^2 +\!252 q^3 +\!441 q^4 +\!1134 q^5+ 1596q^6 \!\cdots ,
$$
\beq
Y_2^{(4)}(\tau )=\sqrt{3}e^{i\pi \tau }\l 1+28q +  126q^2 + 344q^3 + 757q^4 + 1332q^5 + 2198q^6\!\cdots\r ~.
\la{Y4-12}
\eeq
We could obtain these results by extracting the doublet component from the product ($Y\!\!\!\cdot \!\!Y)_2$,
\beq
(Y\!\!\!\cdot \!\!Y)_2=\l Y_2^2-Y_1^2,~2Y_1Y_2\r ,
\la{YxY}
\eeq
 and then by applying in (\ref{YxY}) the expansions given in (\ref{Y12}).

Since we have only two independent holomorphic doublets $Y$ and $(Y\!\!\cdot \!\!Y)_2$, the holomorphic doublet of weight
$k=2n$ is given by
 \beq
(Y^n)_2=a^{(2n-2)}Y+b^{(2n-4)}(Y\!\!\!\cdot \!\!Y)_2 ~,
\la{nYY}
\eeq
 with $a^{(2n-2)}$ and $b^{(2n-4)}$ being the holomorphic forms of weights $(2n-2)$ and $(2n-4)$, respectively.
 They are given by the polynomials $\sum_{m,n}c_{mn} E_4^mE_6^n$ of non-negative powers of $E_4$ and $E_6$.
 Thus, with only  holomorphic forms remaining, 
 we have only two independent doublets. That is why SUSY model building is very constrained and allows us to have some predictions.

Without sticking on holomorphy, from the $Y $ we can have five independent doublets,
$$
Y=(Y_1, Y_2),~~~~Y^*=(Y_1^*, Y_2^*),~~~~(Y\!\!\cdot \!\!Y)_{\bf 2}=(Y_2^2-Y_1^2, 2Y_1Y_2),
$$
\beq
 (Y\!\!\cdot \!\!Y^*)_{\bf 2}=(|Y_2|^2-|Y_1|^2, Y_1Y_2^*+Y_1^*Y_2) ,~~~(Y^*\!\!\cdot \!\!Y^*)_{\bf 2}=(Y_2^{*2}-Y_1^{*2}, 2Y_1^*Y_2^*).
\la{5-dublets}
\eeq
All these have right transformation properties under $S_3$, because the
 matrices $\rho_{[2]}$  [given in Eq. (\ref{rho2-TS})]  are real.
Doublets built from the powers of $Y, Y^*$  higher than $2$  will not be independent.
For example, using the rules (\ref{2by2}) and (\ref{1prby2}), one can verify that
 $\l (Y\!\!\cdot \!\!Y)_{\bf 2}Y^*\r_{\bf 2}=(Y\!\cdot \!Y^*)_{\bf 1}Y\oplus ((Y\!\cdot \!Y^*)_{\bf 1'}Y)_{\bf 2}.$
 From these, and with the help of the modular forms $\tl E_2, E_{4,6}$, and $\ka = (i\tau - i\bar{\tau})$ [given in Eq. (\ref{kapa-transf})],
 one can construct five doublets of any fixed even weight $k$
 and they are given in (\ref{nonhol-Ds}), where $f(\tau)$'s are the corresponding singlet modular forms given in (\ref{fk-from-kaEs}).

Applying (\ref{nonhol-Ds}) we can write down the five doublets with weight $k=2$,
$$
\hs{-4.5cm} k=2~{\rm ~weight~doublets}:~~~~Y,~~~\kappa^2(\tl{E}_2^2\!\!+\!c_1 E_4)Y^* ,~~~\kappa^2\tl{E}_2^*(Y\!\!\cdot \!\!Y)_{\bf 2} ,
$$
 \beq
 \hs{2cm}\kappa^2\tl{E}_2(Y\!\!\cdot \!\!Y^*)_{\bf 2},~~~
 \kappa^4(\tl{E}_2^3\!\!+\!c_2\tl{E}_2E_4\!\!+\!c_3E_6)(Y^*\!\!\cdot \!\!Y^*)_{\bf 2}~ ,
 \la{k2-5doubl}
 \eeq
where $c_{1,2,3}$ are couplings used to construct the given weight form from specific superpositions.
In (\ref{k2-5doubl}) we kept minimal possible powers of the factor $\ka $ [given in Eq. (\ref{kapa-transf})].

Having constructed the doublets of different weights, using the multiplication rule (\ref{2by2}), we can construct the
pseudosinglet modular form with weight $k$, which is given in (\ref{1prime-k}).

Before moving to the next subsection, several comments are in order. 
While Eq. (\ref{fk-from-kaEs}) provides even-weight modular forms with the transformation 
properties given in (\ref{fk}), one can obtain a harmonic Maass form of weight $k$ \cite{BK-rev} 
by imposing the additional constraint $\Delta_k f^{(k)}(\tau) = 0$. 
Here, $\Delta_k = -4y^2 \frac{\partial}{\partial \tau} y^k \frac{\partial}{\partial \bar{\tau}}$ 
represents the weight-$k$ hyperbolic Laplacian.
The form $\tilde{E}_2$ is the weight-2 Maass form captured by Eq. (\ref{fk-from-kaEs}). 
Furthermore, because any holomorphic modular form is a harmonic Maass form, those with 
weights $k > 2$ are simply holomorphic forms. Given that combinations such as $E_4^m E_6^n$ 
are included in Eq. (\ref{fk-from-kaEs}), it follows that all even-weight harmonic 
Maass forms for $k \geq 2$ are subsets of (\ref{fk-from-kaEs}).
If we allow negative powers of $\kappa, \tilde{E}_2$, and $E_{4, 6}$,  even-weight 
Maass forms with $k < 2$ (including negative weights) may emerge from Eq. (\ref{fk-from-kaEs}), 
provided the couplings $C_{lmnp}^{\bar{l}\bar{m}\bar{n}}$ are properly selected. Indeed, 
the holomorphic parts of these negative even-weight Maass forms are$\sim 1/(E_4^m E_6^n)$ \cite{BK}.
Since the present work focuses on models in the vicinity of fixed points, we have excluded 
negative powers of the forms $\tilde{E}_2$ and $E_{4, 6}$ to avoid introducing poles into the 
modular forms. Furthermore, beyond modular symmetry itself, we find no compelling motivation 
for imposing the additional harmonic constraint $\Delta_k f^{(k)}(\tau) = 0$ in our 
construction. While models utilizing harmonic Maass forms are interesting and were 
constructed in \cite{Qu:2024rns}, 
such frameworks would require extra caution when building models near fixed points due to the 
potential emergence of poles.

\subsection{Representations at $\tau =i\infty $ Fixed Point}
\label{ap-Rep-iinf}

In this case we can take the limit $\lan y\ran  \to \infty$ and examine the behavior of the modular forms. From (\ref{E2tl})-(\ref{E4-E6}) we see that
$\tl E_2=E_4=E_6=1$.
Also, with (\ref{Y12}) we see the $Y_2 \to 0$, and therefore, through (\ref{nonhol-Ds}), one can verify that the second component of any modular doublet vanishes, $D_2^{(k)}(i\infty)=0$. On the other hand, using (\ref{1prime-k}), one can see that any pseudosinglet modular form is also zero.
 Therefore, for the doublet and the pseudosinglet modular forms we will have the structures given in (\ref{D1pr-inft}).

For model building, instead of taking $\langle y\rangle$ to be infinite, one can obtain quite good approximations even with $\langle y\rangle \ge 2$. This situation can be referred to as the vicinity of the fixed point $\tau = i\infty$. Since all doublet modular forms (\ref{nonhol-Ds}) are built either from one or two powers of $Y$, with the help of Eqs. (\ref{Y12}) and (\ref{Y4-12}), it is easy to check that, with a good approximation, one can obtain

\beq
{\rm for}~y\stackrel{>}{_\sim }2:~~~\left | \fr{\lan D_2\ran }{\lan D_1\ran}\right |\simeq 8\sqrt{3}e^{-\pi y}~~~{\rm or}~~~~ 16\sqrt{3}e^{-\pi y}.
\la{DevsD1}
\eeq
As a consequence, in the vicinity of the fixed point $\tau = i\infty$, the modular forms can be parametrized as in Eq. (\ref{D1pr-y}).

\subsection{Representations at $\tau =i $ Fixed Point}
\label{ap-Rep-i}

At the fixed point $\tau =i $, among the three modular forms $\tl E_2, E_{4}$, and $E_6$, only $E_4$ has a nonvanishing value,
\beq
{\rm at} ~\tau=i:~~~~\tl E_2=E_6=0 ,~~~~E_4\simeq 1.4558.
\la{tau-i-E246}
\eeq
Therefore, in the construction of the singlet modular forms (\ref{fk-from-kaEs}), only $E_4$ and $\ka $ will participate, and we will have
\beq
{\rm at} ~\tau=i:~~~~f^{(k)}=\sum C_{m}^{\ov m}
 \ka^{2p}E_4^{\hs{0.03cm} m} (E_4^{\hs{0.05cm}\ov m})^*  ,~~~
{\rm with}~~~p=2\ov m,~~~k=4(m-\ov m),
\la{tau-i-fk}
\eeq
from which we can see that nonvanishing  singlet modular forms have weights $4n$, while singlets with  weights $(4n+2)$
vanish,
\beq
{\rm at} ~\tau=i:~~~~ {\bf 1}^{(4n)}\neq 0 ,~~~~{\bf 1}^{(4n+2)}= 0  ,
\la{tau-i-1s}
\eeq
as also shown in Eq. (\ref{1-1pr-i}).

For the modular doublets, it turns out that  for $Y$ and $(Y\!\!\cdot \!\!Y)_2\propto Y^{(4)}$,
 \beq
{\rm at} ~\tau=i:~~~~ Y\propto \l \!\!\begin{array}{c}
                                 1 \\
                                 \fr{1}{\sq{3}}
                               \end{array}\!\!\r ,~~~~~
                              (Y\!\cdot \!Y)_2\propto \l \!\!\begin{array}{c}
                                 1 \\
                                 -\sq{3}
                               \end{array}\!\!\r ,
\la{tau-i-YY4}
\eeq
both are real. Using (\ref{tau-i-1s}) and (\ref{tau-i-YY4}) in (\ref{nonhol-Ds}),  the modular doublets with the weights
$(4n+2)$ and $4n$ yield the structures shown in Eq. (\ref{Ds-at-i}).

Finally, with the forms of the doublets in (\ref{Ds-at-i}) and using (\ref{1prime-k}), the pseudosinglet 
modular forms of weights $(4n+2)$ and $4n$, respectively, yield
\beq
{\rm at} ~\tau=i:~~~~ {\bf (1')}^{(4n+2)}\neq 0 ,~~~~{\bf (1')}^{(4n)}= 0 ,
\la{tau-i-1prs}
\eeq
which is also indicated in Eq. (\ref{1-1pr-i}).

\subsection{Representations at $\tau =\om $ Fixed Point}
\label{ap-Rep-om}

At the fixed point $\tau =\om $, the modular forms $\tl E_2, E_{4}$ and $E_6$ take values
\beq
{\rm at} ~\tau=\om :~~~~\tl E_2=E_4=0 ,~~~~E_6\simeq 2.8815.
\la{tau-om-E246}
\eeq
Therefore, in the construction of the singlet modular forms (\ref{fk-from-kaEs}) only $E_6$ and $\ka $ will participate
as
\beq
{\rm at} ~\tau=\om:~~~~f^{(k)}=\sum C_{m}^{\ov m}
 \ka^{2p} E_6^{\hs{0.03cm} m} (E_6^{\hs{0.05cm}\ov m})^* ~~
{\rm with}~~~p=3\ov m,~~~k=6(m-\ov m).
\la{tau-om-fk}
\eeq
From (\ref{tau-om-fk})  we can see that nonvanishing  singlet modular forms have weights $6n$, while singlets with  weights $(6n+2)$ and $(6n+4)$
vanish, as given in Eq. (\ref{1-1pr-om}).

For the modular doublets, it turns out that  $Y$ and $(Y\!\!\cdot \!\!Y)_2\propto Y^{(4)}$,
 \beq
{\rm at} ~\tau=\om:~~~~ Y\propto \l \!\!\begin{array}{c}
                                 1 \\
                                 -i
                               \end{array}\!\!\r ,~~~~~
                              (Y\!\cdot \!Y)_2\propto \l \!\!\begin{array}{c}
                                 1 \\
                                 i
                               \end{array}\!\!\r \propto Y^*,
\la{tau-om-YY4}
\eeq
have structures of the conjugate of each other. Using,  (\ref{tau-om-YY4}) and (\ref{1-1pr-om}) in (\ref{nonhol-Ds}),  the modular doublets with the weights
$(6n+p)$ (where $p=0, 2, 4$)  have the structures given in Eq. (\ref{Ds-at-om}).

Finally, with the forms of the modular doublets in (\ref{Ds-at-om}) and using (\ref{1prime-k}), 
at $\tau =\om $ we can obtain the different weight  pseudosinglet modular forms,  
which are given in Eq.  (\ref{1-1pr-om}).

\renewcommand{\theequation}{B.\arabic{equation}}\setcounter{equation}{0}

\section{Kinetic Couplings and Fixing $\lan \tau \ran$}
\la{app-kin}


For the fermionic state $\psi $, with the weight $k_{\psi}$, using the two-component Wyel spinor, the kinetic Lagrangian
invariant under the modular transformation
$$
\psi'=(c\tau+d)^{k_{\psi }}\psi ~,~~~~~~\bar{\psi}'=(c\bar{\tau}+d)^{k_{\psi }}\bar{\psi }~,
$$
has the form
\beq
{\cal L}_{kin}(\psi)=\fr{i}{2}\l \pl_{\mu}\bar{\psi}\bar{\si }^{\mu}\psi -\bar{\psi}\bar{\si }^{\mu}\pl_{\mu}\psi\r (i\bar{\tau}-i\tau )^{k_{\psi }}
-\fr{k_{\psi}}{2}\l \pl_{\mu}\bar{\tau}+\pl_{\mu}\tau \r (i\bar{\tau}-i\tau )^{k_{\psi }-1}\bar{\psi}\bar{\si }^{\mu}\psi ~.
\la{Lk-psi}
\eeq
Note that, we are limiting ourselves by focusing solely on the minimal kinetic couplings,
without accounting for kinetic mixings between different flavors.
With the vacuum expectation value of the $\lan \tau \ran $, the canonically normalized 
fermion state $\psi_c$ will be related to $\psi $ as follows:
\beq
\psi =\fr{1}{(i\lan \bar{\tau }\ran-i\lan \tau \ran )^{k_{\psi}/2}}\psi_c ~.
\la{canon-ferm}
\eeq
The kinetic term for the field $\tau $ is
\beq
{\cal L}_{kin}(\tau)=\fr{\La^2}{(i\bar{\tau}-i\tau )^2}\pl_{\mu}\bar{\tau }\pl^{\mu}\tau ,
\la{Lk-tau}
\eeq
which is invariant under (\ref{tau-to-tau1}) transformation. The $\La $ is some mass scale. The canonically normalized state
$\tau_0=\fr{1}{\sqrt{2}}(x_0+iy_0)$ is
related to $\tau $ as
\beq
\tau =\lan \tau \ran +i\fr{\lan \bar{\tau }\ran \!-\!\lan \tau \ran }{\sqrt{2}\La }(x_0+iy_0)+\cdots ,
\la{tau-can}
\eeq
where $``..."$ stand for the higher powers of the fields, which are not relevant to us.

Near the $\tau =i\infty $ and $\tau=i$ fixed points,
the desired values of $\lan \tau \ran $ can be obtained in a rather simple way via the two invariants,
\beq
X_1=\ka^2|\tl E_2|^2~,~~~X_2=\ka^4 \l \tl E_2^2E_4^* +\tl E_2^{*2}E_4\r ,
\la{X12-invs}
\eeq
through which the potential for the $\tau $ can be constructed,
\beq
V(\tau )=\fr{\lam_1}{2}\Lambda^4(X_1-a_1)^2+\fr{\lam_2}{2}\Lambda^4(X_2-a_2)^2 .
\la{Vtau}
\eeq
With the proper choice of the parameters $\lam_{1,2}$ and $a_{1,2}$  we get the desirable value of the $\lan \tau \ran $.
Below, for the models near the $\tau =i\infty $ and $\tau =i$ fixed points,
 we give the selections and obtained values of $\lan \tau \ran $, which correspond to the minimum of the potential  $V(\tau )$:
$$
{\rm Near} ~~~\tau =i\infty  :~~ \lan \tau \ran =4.5i+0.01 ,~~\{\lam_1,\lam_2, a_1, a_2\}\!=\!\{ 10^{-5},~ 10^{-7},~12.568,~508.99\},~   ~
$$
\beq
\{M_{x_0}, M_{y_0}\}\!\simeq \!\La \{8.4\cdot 10^{-12}, 1\}  ,
\la{select-pars1}
\eeq
$$
{\rm Near} ~~~\tau =i  :~~ \lan \tau \ran =i(1\!+\!10^{-5})\!-\!5\cdot 10^{-7} ,~~
\{\lam_1,\lam_2, a_1,a_2\}\!=\!\{1, 0.02,~1.54\cdot 10^{-10},~4.47\cdot 10^{-10}\},~   ~
$$
\beq
\{M_{x_0}, M_{y_0}\}\!\simeq \!\La \{8.7\cdot 10^{-8}, 4.7\cdot 10^{-5}\}.
\la{select-pars2}
\eeq
In (\ref{select-pars1}) and (\ref{select-pars2}) we also indicated the masses  of the canonically normalized states  ${x_0, y_0}$ [defined in (\ref{tau-can})].
 As we see, for the $y=4.5$
[case of Eq.  (\ref{select-pars1})] with
 $\La \simeq M_{Pl}=1.22\cdot 10^{19}$GeV,  the mass of $x_0$ is$\simeq 10^{8}$GeV.  This value is completely sufficient
 for suppressing the flavor-violating processes because the couplings of $x_0$  with the matter have additional suppression
 factor $\sim e^{-\pi \lan y\ran} (\sim 10^{-6}$
for $\lan y\ran =4.5$). For the selection given in  (\ref{select-pars2}), corresponding to the model near the $\tau =i$ fixed point,
both states $x_0$ and $y_0$ have sufficiently large masses($\stackrel{>}{_\sim }\!10^{12}$GeV for $\La \simeq M_{Pl}$).

Near the $\tau=\om $ fixed point,
the desired values of $\lan \tau \ran $ can be obtained via the potential
$$
V_{\om}(\tau )=\fr{\lam_3}{2}\Lambda^4(X_3-a_3)^2+\fr{\lam_4}{2}\Lambda^4(X_4-a_4)^2 ,
$$
\beq
{\rm with:}~~~X_3=\ka^6|E_6|^2~,~~~X_4=\ka^6 \l \tl E_2E_4E_6^* +\tl E_2^{*}E_4^*E_6\r .
\la{Vom-pot}
\eeq
The minimum of the potential $V_{\om}(\tau )$ 
near the $\tau =\om $  fixed point is obtained for the selection
$$
{\rm Near} ~~\tau \!=\!\om   :~~ \lan \tau \ran \!=\!\om +(0.5196+i1.093)\tm 10^{-5} ,
~~\{\lam_3, \lam_4, a_3, a_4\}\!=\!\{ 0.1304,~ 0.02,~3.503,~1.368\tm 10^{-9}\}, 
$$
\beq
\{M_{x_0}, M_{y_0}\}\!\simeq \!\La \{1.15\cdot 10^{-6}, 8.46\cdot 10^{-5}\}.
\la{select-pars-om}
\eeq
We see that for $\La \simeq M_{Pl}$ both states $x_0$ and $y_0$ are pretty 
heavy($\approx \!1.4\cdot 10^{13}$GeV and $10^{15}$GeV, respectively).

\renewcommand{\theequation}{C.\arabic{equation}}\setcounter{equation}{0}

\section{Neutrino Parametrization}
\la{ApU}


In a standard parametrization \cite{ParticleDataGroup:2024cfk}, the lepton mixing matrix $U$ has the following form:
\begin{equation}
U=
\left(\begin{array}{ccc}c_{13}c_{12}&c_{13}s_{12}&s_{13}e^{-i\delta}\\
-c_{23}s_{12}-s_{23}s_{13}c_{12}e^{i\delta}&c_{23}c_{12}-s_{23}s_{13}s_{12}e^{i\delta}&s_{23}c_{13}\\
s_{23}s_{12}-c_{23}s_{13}c_{12}e^{i\delta}&-s_{23}c_{12}-c_{23}s_{13}s_{12}e^{i\delta}&c_{23}c_{13}
\end{array}\right),
\la{Ulept}
\end{equation}
with $s_{ij}=\sin\theta_{ij}$ and $c_{ij}=\cos\theta_{ij}$. The phase matrices $P$ and $P'$ are given by
\beq
P={\rm Diag}\l e^{i\om_1}~,~e^{i\om_2}~,~e^{i\om_3}\r,~~~
P'={\rm Diag}\l 1~,~e^{i\rho_1}~,~e^{i\rho_2}\r,
\la{Ps}
\eeq
where $\om_{1,2,3}$ are some phases and $\rho_{1,2}$ are Majorana phases appearing in the neutrinoless double-$\bt $ decay amplitude.

The relation (\ref{barMnu}) allows us to express the entries of the $M_{\nu }$ in terms of the $\om_i$ phases and ${\cal A}_{ij}$, where
\beq
{\cal A}_{ij}=(U^*P'M_{\nu}^{\rm Diag}U^{\dag })_{ij} .
\la{defA}
\eeq

For instance, for the neutrino scenario, presented in Sec.  \ref{sec-inf-nu},
using the mass matrix (\ref{Z2T-nu-matrix}), we can express
 $\bar m, a, b, \ep $, and  $\ep_1$ as follows:
 $$
\bar m=e^{2i\om_3}{\cal A}_{33}, ~~~a=e^{i(\om_2-\om_3 )}\fr{{\cal A}_{23}}{{\cal A}_{33}},~~~
b=e^{2i(\om_1-\om_3 )}\fr{{\cal A}_{11}}{{\cal A}_{33}},
$$
\beq
\ep =e^{i(\om_1-\om_3 )}\fr{{\cal A}_{13}}{{\cal A}_{33}},~~~~
\ep_1=e^{i(\om_1-\om_3 )}\l \fr{{\cal A}_{12}}{{\cal A}_{23}}-\fr{{\cal A}_{13}}{{\cal A}_{33}}\r .
\la{nu-pars-A}
\eeq
On the other hand, from (\ref{ab-ab0}) we can express the original parameters as
$$
\tan\te_e=\left | \fr{a\ep_1}{b-\ep^2}\right | ,~~~\eta =-Arg\l \fr{a\ep_1 }{\ep^2-b}\r ,~~~a_0=ac_e+\ep s_ee^{-i\eta },
$$
\beq
b_0=\pm \fr{1}{c_e\sqrt{n}}\left [ b\!-\!\ep^2\!+\!c_e^2(\ep c_e\!-\!as_ee^{i\eta })^2\right ]^{1/2} ,~~~
\ep_0=\pm c_e\sqrt{n}\fr{\ep c_e-as_ee^{i\eta }}{\left [ b\!-\!\ep^2\!+\!c_e^2(\ep c_e\!-\!as_ee^{i\eta })^2\right ]^{1/2}} ~.
\la{orig-pars}
\eeq
%
Using (\ref{nu-pars-A}) in (\ref{orig-pars})
for the parameter $\te_e$, for IO and NO cases, respectively,  we find the  results given in Eq. (\ref{tee-values}).

\bibliographystyle{unsrt}

\begin{thebibliography}{99}


\bibitem{Capozzi:2021fjo}
F.~Capozzi, E.~Di Valentino, E.~Lisi, A.~Marrone, A.~Melchiorri and A.~Palazzo,
Phys. Rev. D \textbf{104},  083031 (2021);
%
M.~C.~Gonzalez-Garcia, M.~Maltoni and T.~Schwetz,
Universe \textbf{7},  459 (2021).


\bibitem{Froggatt:1978nt}
  C.~D.~Froggatt and H.~B.~Nielsen,
   Nucl.\ Phys.\ B{\bf 147}, 277 (1979).




\bibitem{Dudas:1995yu}
  E.~Dudas, S.~Pokorski, and C.~A.~Savoy,
   Phys.\ Lett.\ B {\bf 356}, 45 (1995).

\bibitem{Chen:2008tc}
  M.~-C.~Chen, D.~R.~T.~Jones, A.~Rajaraman, and H.~-B.~Yu,
   Phys.\ Rev.\ D {\bf 78}, 015019 (2008).



\bibitem{Tavartkiladze:2011ex}
  Z.~Tavartkiladze,
  Phys.\ Lett.\  B {\bf 706}, 398 (2012);
%
%
Phys. Rev. D \textbf{87}, 075026 (2013).
%
Z.~Tavartkiladze,
Phys. Rev. D \textbf{106}, no.11, 115002 (2022).


%
%

\bibitem{Pakvasa:1977in}
S.~Pakvasa and H.~Sugawara,
Phys. Lett. B \textbf{73}, 61-64 (1978).





%
%


%
%

\bibitem{deAdelhartToorop:2011re}
R.~de Adelhart Toorop, F.~Feruglio and C.~Hagedorn,
Nucl. Phys. B \textbf{858}, 437-467 (2012).

\bibitem{Feruglio:2017spp}
F.~Feruglio,
[arXiv:1706.08749 [hep-ph]].


\bibitem{Kobayashi:2018vbk}
T.~Kobayashi, K.~Tanaka and T.~H.~Tatsuishi,
Phys. Rev. D \textbf{98}, no.1, 016004 (2018).

\bibitem{Kobayashi:2019rzp}
T.~Kobayashi, Y.~Shimizu, K.~Takagi, M.~Tanimoto and T.~H.~Tatsuishi,
PTEP \textbf{2020}, no.5, 053B05 (2020).



\bibitem{Du:2020ylx}
X.~Du and F.~Wang,
JHEP \textbf{02}, 221 (2021).



\bibitem{Novichkov:2018ovf}
P.~P.~Novichkov, J.~T.~Penedo, S.~T.~Petcov and A.~V.~Titov,
JHEP \textbf{04}, 005 (2019).

\bibitem{Ding:2019gof}
G.~J.~Ding, S.~F.~King, X.~G.~Liu and J.~N.~Lu,
JHEP \textbf{12}, 030 (2019).

\bibitem{Okada:2020ukr}
H.~Okada and M.~Tanimoto,
Phys. Rev. D \textbf{103}, no.1, 015005 (2021).



\bibitem{Feruglio:2021dte}
F.~Feruglio, V.~Gherardi, A.~Romanino and A.~Titov,
JHEP \textbf{05}, 242 (2021);
F.~Feruglio,
Phys. Rev. Lett. \textbf{130}, no.10, 101801 (2023).



\bibitem{Kobayashi:2021pav}
T.~Kobayashi, H.~Otsuka, M.~Tanimoto and K.~Yamamoto,
Phys. Rev. D \textbf{105}, no.5, 055022 (2022).

\bibitem{Kikuchi:2023cap}
S.~Kikuchi, T.~Kobayashi, K.~Nasu, S.~Takada and H.~Uchida,
Phys. Rev. D \textbf{107}, no.5, 055014 (2023).

\bibitem{Feruglio:2023mii}
F.~Feruglio,
JHEP \textbf{03}, 236 (2023).

\bibitem{Meloni:2023aru}
D.~Meloni and M.~Parriciatu,
JHEP \textbf{09}, 043 (2023).





\bibitem{Marciano:2024nwm}
S.~Marciano, D.~Meloni and M.~Parriciatu,
JHEP \textbf{05}, 020 (2024).

\bibitem{Nomura:2024ouj}
T.~Nomura, M.~Tanimoto and X.~Y.~Wang,
Eur. Phys. J. C \textbf{84}, no.12, 1329 (2024).

\bibitem{Kobayashi:2023zzc}
For reviews and references see:\\
T.~Kobayashi and M.~Tanimoto,
Int. J. Mod. Phys. A \textbf{39}, no.09n10, 2441012 (2024);\\
%
G.~J.~Ding and S.~F.~King,
Rept. Prog. Phys. \textbf{87}, no.8, 084201 (2024).


\bibitem{Kumar:2023moh}
R.~Kumar, P.~Mishra, M.~K.~Behera, R.~Mohanta and R.~Srivastava,
Phys. Lett. B \textbf{853}, 138635 (2024).



\bibitem{Granelli:2025lds}
A.~Granelli, D.~Meloni, M.~Parriciatu, J.~T.~Penedo and S.~T.~Petcov,
[arXiv:2505.21405 [hep-ph]].



\bibitem{Qu:2024rns}
B.~Y.~Qu and G.~J.~Ding,
JHEP \textbf{08}, 136 (2024).



\bibitem{Qu:2025ddz}
B.~Y.~Qu, J.~N.~Lu and G.~J.~Ding,
JHEP \textbf{11}, 140 (2025).

%
%
%
%

\bibitem{kaneko}
M. Kaneko and D. Zagier, A generalized Jacobi theta function and quasimodular forms,
The moduli space of curves (Texel Island, 1994), Progr. Math., vol. 129, Birkhauser
Boston, Boston, MA, 1995, pp. 165-172. MR1363056 (96m:11030).

\bibitem{apostol}
T. M. Apostol, Modular functions and Dirichlet series in number theory,
$2^{\rm nd}$ ed., 1990 Springer-Verlag New York, Inc.


\bibitem{zagier}
D Zagier, ``Elliptic modular forms and their applications'',  The 1-2-3 of modular forms: Lectures at a summer school in Nordfjordeid,
Norway, 2008 -- Springer.



\bibitem{Nagatomo} K. Nagatomo, Y. Sakai and D. Zagier,   ``Modular Linear Differential Operators and Generalized Rankin-Cohen Brackets'',
  [arXiv:2210.10686 [math.NT]]. 
 



\bibitem{deMedeirosVarzielas:2025byb}
I.~de Medeiros Varzielas, M.~S.~Liu, A.~Sengupta and J.~Talbert,
[arXiv:2512.19789 [hep-ph]].


\bibitem{in-prep}
Z. Tavartkiladze (to be published).


\bibitem{Shafi:2000su}
Q.~Shafi and Z.~Tavartkiladze,
Phys. Lett. B \textbf{482}, 145-149 (2000).



\bibitem{ParticleDataGroup:2024cfk}
S.~Navas \textit{et al.} [Particle Data Group],
Phys. Rev. D \textbf{110}, no.3, 030001 (2024).


\bibitem{BK-rev}
For a review see:
K. Bringmann and S. Kudla, ``A classification of harmonic Maass forms'',
[arXiv:1609.06999 [math.NT]]; and references therein.


\bibitem{BK}
K Bringmann and B Kane, ``Ramanujan-like formulas for Fourier coefficients of all meromorphic cusp forms'',
[arXiv:1603.09250 [math.NT]].


\end{thebibliography}

\end{document}